# Multiscale autonomous forecasting of plasma systems' dynamics using neural networks

F. Faraji*[1] and M. Reza*

* Plasma Propulsion Laboratory, Department of Aeronautics, Imperial College London, London, United Kingdom

**Abstract**: Plasma systems exhibit complex multiscale dynamics, resolving which poses significant challenges for conventional numerical simulations. Machine learning (ML) offers an alternative by learning data-driven representations of these dynamics. Yet existing ML time-stepping models suffer from error accumulation, instability, and limited long-term forecasting horizons. This paper demonstrates the application of a hierarchical multiscale neural network architecture for autonomous plasma forecasting. The framework integrates multiple neural networks trained across different temporal scales to capture both fine-scale and large-scale behaviors while mitigating compounding error in recursive evaluation. By structuring the model as a hierarchy of sub-networks, each trained at a distinct time resolution, the approach effectively balances short-term resolution with long-term stability. Fine-scale networks accurately resolve fast-evolving features, while coarse-scale networks provide broader temporal context, reducing the frequency of recursive updates and limiting the accumulation of small prediction errors over time. We first evaluate the method using canonical nonlinear dynamical systems and compare its performance against classical single-scale neural networks. The results demonstrate that single-scale neural networks experience rapid divergence due to recursive error accumulation, whereas the multiscale approach improves stability and extends prediction horizons. Next, our ML model is applied to two plasma configurations of high scientific and applied significance, demonstrating its ability to preserve spatial structures and capture multiscale plasma dynamics. By leveraging multiple time-stepping resolutions, the applied framework is shown to outperform conventional single-scale networks for the studied plasma test cases. Additionally, another great advantage of our approach is its parallelizability by design, which enables the development of computationally efficient forecasters. The results of this work position the hierarchical multiscale neural network as a promising tool for efficient plasma forecasting and digital twin applications.

## Section 1: Introduction

Plasmas are inherently multiscale systems, governed by interactions and phenomena that span several orders of magnitude in both space and time. The multiscale nature poses significant challenges for plasma simulations, as they typically need computationally demanding models that can accurately resolve fine-scale dynamics while capturing large-scale collective behaviors [1]-[4]. This also presents a challenge for the development of machine learning (ML) models for plasma systems, in that it requires time-stepping architectures that can efficiently learn and generalize across a broad spectrum of scales.

Time-stepping refers to the iterative prediction of a system's state at future times based on current and past states. Traditional equation-based time-steppers have long been the backbone of scientific computing in fields such as computational fluid dynamics, plasma simulation, and climate modelling. These methods rely on discretizing differential equations that govern the underlying physical processes and then integrating the discretized set of equations using numerical techniques such as the Euler method [5], Runge-Kutta methods [6], and implicit schemes like Crank-Nicolson [7]. These classical time-stepping methods are known for their robustness, stability and well-understood error bounds based on numerical analysis principles.

In cases where the governing equations are unknown or partially understood, ML approaches provide a potential alternative for learning complex, data-driven representations of system dynamics. The use of ML architectures as time-stepping models in scientific computing has gained significant attention [8]-[16]. Several ML architectures have been employed for time-stepping, each with inherent challenges and specific limitations in autonomous recursive time-stepping.

Recurrent Neural Networks (RNNs), including Long Short-Term Memory (LSTM) [17] and Gated Recurrent Units (GRUs) [18], have been extensively used for time-series prediction due to their ability to model temporal dependencies [19]-[23], showing promise in forecasting and sequence prediction tasks. Their key strengths include memory retention of past states and the ability to capture long-term dependencies. However, they are susceptible to vanishing and exploding gradients and accumulate prediction errors in medium to long-term forecasts.

---

[1] **Corresponding Author** (f.faraji20@imperial.ac.uk)

Transformers, originally developed for natural language processing, have emerged as powerful time-series predictors due to their attention mechanism [24]-[26], which enables them to capture long-term dependencies, making them ideal for problems where past states influence future predictions. However, unless trained on substantially wide datasets, they are prone to overfitting, especially when dealing with limited scientific datasets.

Physics-Informed Neural Networks (PINNs) [27]-[29] incorporate physical laws as loss functions, promoting physically consistent predictions. Their strengths include the integration of domain knowledge and improved generalization for physics-constrained tasks. A notable challenge with PINNs is their underlying assumption that the system of equations governing a system's evolution is fully known, allowing these equations to be incorporated into the loss function [30].

Neural networks can be used to learn discretized differential equations describing the dynamics through capturing the flow maps. Flow map is a mathematical representation that describes how the state of a dynamical system evolves over time. Neural networks, particularly Residual Networks (ResNets) [31], have been instrumental in approximating discrete-time evolution of the dynamical system and flow maps. ResNets are known for their effectiveness in training very deep neural networks through the use of residual connections, which help mitigate the vanishing gradient problem common in deep networks. By introducing skip/shortcut connections that bypass one or more layers, ResNets allow gradients to flow through the network during backpropagation. As a result, in the context of time steppers, instead of learning the direct mapping $F(x)$ from input $x^k$ to output $x^{k+1}$, where $x^{k+1} = x^k + F(x)$, ResNets learn the residual mapping or difference between successive states in time $(H(x) = x^{k+1} - x^k)$ similar to classical Euler integration.

It is important to note that there are two ways by which ML architectures can be applied for forecasting. One is *one-step-ahead* forecasting, where each time step directly predicts the next with a forecast window of a single time step. The other is *recursive* or *closed-loop* (also known as *autonomous* or *generative*) forecasting, which involves advancing forward in time self-consistently by rolling out predictions based solely on previous time step(s) predictions.

While most ML architectures can perform well at single step or finite number of steps prediction, extending forecast window in recursive manner in longer terms face critical challenges related to error propagation and stability [32]. Recursive predictions lead to compounding errors, causing significant divergence from the ground truth after only a few steps. Most of these models also lack inherent numerical stability properties common in classical solvers, leading to instabilities in long-term forecasts.

This challenge is especially pronounced in chaotic systems, where even small deviations are quickly amplified due to their intrinsic sensitivity to initial conditions, causing rapid divergence from the true trajectory. Additionally, the high error sensitivity of these models imposes stringent requirements on the quality of training data, demanding large datasets with minimal noise to minimize errors in the learned model.

In contrast to neural-network time steppers, Dynamic Mode Decomposition (DMD) models [33]-[39] generally do not suffer from the problem of error accumulation caused by recursive feedback. This is because they compute future states in a single step by applying a learned linear operator directly to the initial state. Despite its strengths, DMD has several limitations including inability of predicting transient dynamics in complex systems due to its reliance on temporally fixed modes and eigenvalues representations.

One way to address error accumulation in forecasting ML models is by feeding the model with a few system measurements at each time step. By providing real-time observations, the model can adjust its predictions, effectively resetting accumulated errors and improving overall accuracy, though at the cost of losing self-consistency in predictions. This approach resembles ML-based inference tasks [40]-[44], where the full spatial flow field is inferred from a few local measurements. These measurements serve as anchors, correcting deviations from the predicted trajectory and keeping the model aligned with the true system state.

In the context of self-consistent predictions, however, the hierarchical time-stepping scheme proposed by Liu et al. [45] offers a promising solution for reducing error accumulation in long-term predictions of ML models. By combining neural network time-steppers across multiple temporal scales, the method addresses some of the key challenges of multiscale dynamics, including numerical stiffness and error propagation.

In the hierarchical approach proposed by Liu et al. [45], fine-grained time-steppers ensure short-term accuracy, while coarse-grained models periodically "reset" predictions over longer intervals, limiting cumulative errors. This layered approach balances accuracy with stability, making long-term simulations more reliable.

The hierarchical framework also enhances computational efficiency through parallelization and vectorization, enabling simultaneous processing across temporal scales. This capability leverages modern high-performance computing (HPC) resources for faster simulations compatible with needs such as real-time prediction and control.

The hierarchical time-stepping framework has been demonstrated in Ref. [45] for simple nonlinear systems to demonstrate superior performance compared to single-scale time-stepper components and other approaches such LSTM, Echo State Network (ESN) and Reservoir Computing (RC) [46]. Those results motivated this research toward using the method's potential for scalable and efficient long-term forecasting in more complex dynamical systems, as well as more realistic plasma configurations of applied relevance.

**Section 2: Description of the multiscale neural network architecture**

The hierarchical multiscale neural network model consists of multiple deep learning components arranged in a structured framework to capture the dynamics of a system at different temporal resolutions.

At its core, the method decomposes the system's evolution into a hierarchy of flow maps with different resolutions in time, where each neural network is trained independently to represent discrete-time mapping of the system across its respective time scale. Each of these networks with a specific step size represents a "sub-model". The multiscale architecture then aggregates these individual sub-models to form a multiscale model. Networks operating at larger time steps provide coarse predictions, while finer-scale networks refine these predictions over shorter intervals, allowing for both computational efficiency and detailed resolution of intricate behaviors. As a result, the method can efficiently incorporate multiscale dynamics through its constituent sub-models. A schematic of such hierarchical time stepping scheme is presented in Figure 1.

A key advantage of this hierarchical design is its ability to reduce error accumulation over extended prediction horizons. Coarse time-stepping networks prevent the rapid propagation of compounding errors, while finer networks ensure that short-scale behaviors are resolved. Additionally, the model's structure allows for parallelization, significantly improving computational efficiency when compared to conventional integration methods.

Each neural network within the hierarchy is a residual network composed of multiple layers, which can be, for example, fully connected layers. However, in this work, we primarily utilize LSTM in the residual networks' layers to capture temporal dependencies more effectively. The input to each network is the current state of the system, and the output represents the predicted state increment over its corresponding time step.

The hierarchical coupling of these networks is achieved through a sequential time-stepping process. The process begins with the coarsest-scale network generating an initial approximation of the system's state evolution over extended intervals. These predictions are then refined by progressively finer-scale networks, which fill in the small-scale resolutions.

To maximize computational efficiency, the framework employs a vectorized computation strategy, as implemented by Liu et al. [45], which enables parallel processing of state predictions over time. Instead of sequentially stepping through the entire time sequence, the approach partitions the sequence into smaller sub-sequences (segments), where each segment's length corresponds to the largest time step in the hierarchy (see Figure 1). At the coarsest level, state forecasts serve as initial conditions for the finer-scale networks that enable parallel execution of fine-grained predictions. Hence, the coarse intervals can be stacked together and computed simultaneously. This parallelization is particularly beneficial for high-dimensional dynamical systems, where traditional single scale time-stepping methods become computationally expensive.

For the entirety of this work, we have used the open-source code developed by Liu et al. and available on GitHub [47].

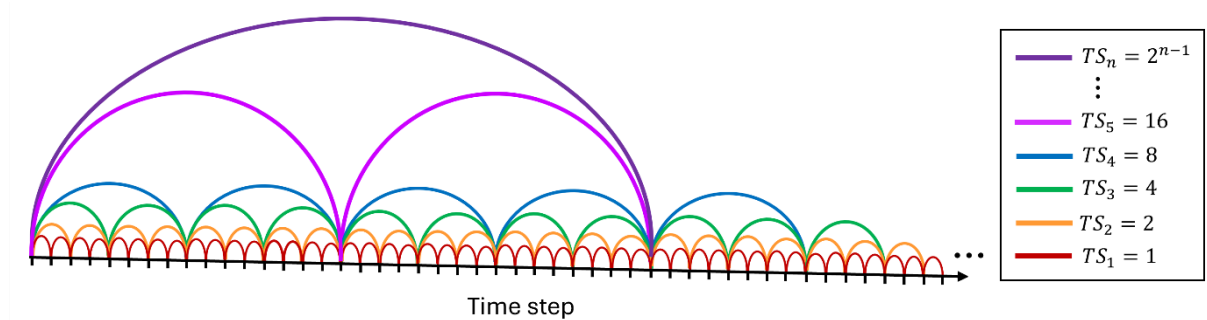

Figure 1: Schematic representation of the hierarchical multiscale time-stepper scheme. Each color represents a "sub-model" with a specific step size denoted by $TS$.

**Section 3: Demonstrative example**

Before proceeding to the results, it is important to illustrate the limitations of applying standard neural networks, such as LSTM or ResNet, recursively over time. To this end, we present a demonstrative example in this section. We utilize a simple canonical dynamical system, the Van der Pol oscillator – a nonlinear system which has limit cycle behavior in phase space, exhibiting periodic dynamics. The details of data generation and the specification of the neural network architecture trained on this data are provided in Appendix A.

We trained multiple neural networks with different step sizes to investigate their forecasting performance on this example. Specifically, we trained ResNet architectures with fully connected (FC) layers as well as ResNet with LSTM-based networks with step sizes of $1, 2, 4, \ldots, 4096$. Each of these models are trained to predict future states either in a one-step-ahead or multi-step-ahead recursive mode, allowing us to explore the impact of network architecture and different training strategies on the model's ability to capture the system's dynamics. Additionally, we implemented the hierarchical multiscale approach, where forecasts from networks trained on different time scales were aggregated to enhance long-term predictive performance.

When a model is employed in a recursive (closed-loop) manner, i.e., its output at each time step serves as input for the next, standard neural networks such as single-step-size ResNet and LSTM exhibit significant error accumulation over time, resulting in significant deterioration of the prediction accuracy. This issue arises due to the compounding of prediction errors at each time step. Unlike one-step predictions, where the neural network corrects for immediate inaccuracies by receiving ground-truth inputs at each time step, recursive application relies on its previous outputs, which are increasingly prone to deviation from the true underlying sequence. As the model progresses through time steps, even minor prediction errors accumulate, amplifying deviations and leading to substantial inaccuracies. The forecasting results of the individual networks with different step sizes in Figure 2 illustrate this effect. Initially, the forecasts align well with the ground truth, but as time progresses, deviations grow, particularly in models trained using smaller step sizes, where recursive evaluations are more frequent.

This issue exists not only for predicting sequences beyond the training data but also in generating the training sequence itself in recursive mode. Additionally, while the presence of chaos in chaotic systems and transient phenomena in non-periodic systems both limit the forecasting window (by increasing sensitivity to errors in previous time steps and reducing the generalizability of the learned model beyond the training data, respectively), the presented example shows that challenge persists even in fully periodic systems.

The root cause of this issue lies in the mismatch between training and autonomous sequence generation and forecasting strategy. During training, ML models are typically exposed to ground-truth sequences, receiving accurate inputs at each time step. However, during forecasting in a recursive setup, the model must rely on its own predictions, creating a feedback loop of errors.

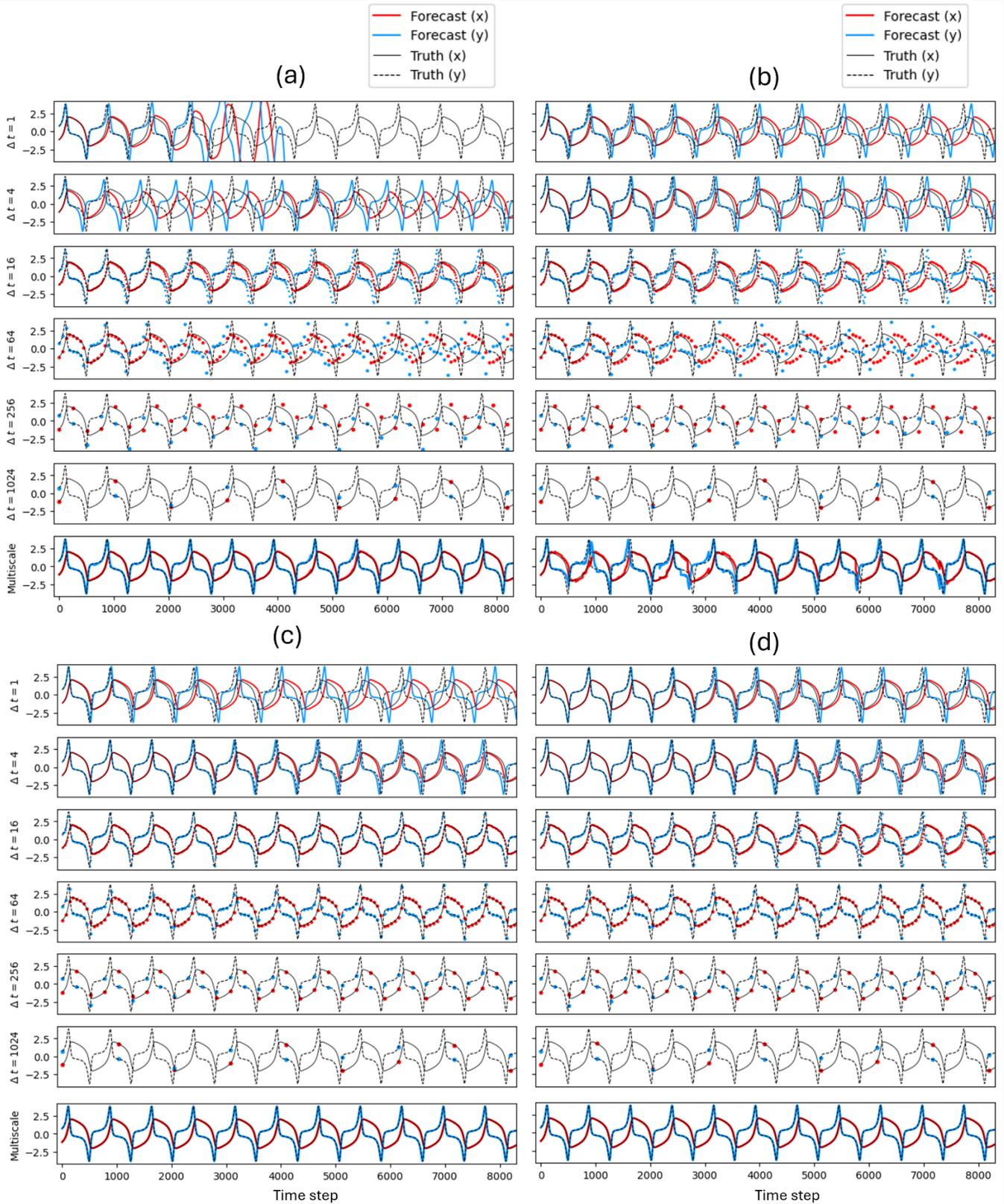

Figure 2: Forecasts from different neural network architectures trained using varying step sizes compared against ground-truth trajectories for the Van der Pol oscillator: FC ResNet trained using **(a)** 1-step-ahead and **(b)** 4-step-ahead training strategy, and LSTM-based ResNet trained using **(c)** 1-step-ahead and **(d)** 4-step-ahead training strategy. Each row corresponds to a model trained using a specific step size, with the bottom-most row in each panel presenting the multiscale forecast, incorporating predictions from individual networks trained at different step sizes.

To mitigate this mismatch, we explored "multi-step-ahead" training strategy, where the model learns to predict multiple time steps into the future by recursively using its own predictions during training over a fixed horizon window. This approach, also known as "recursive sequence training", forces the model to rely on its own predictions for subsequent time steps while being trained, improving the long-term forecasting ability of the model.

|  | Loss [%] | |
| --- | --- | --- |
|  | FC | LSTM |
| $n_f = 1$ | 3.245 | 0.294 |
| $n_f = 4$ | 14.55 | 0.101 |
| $n_f = 8$ | 14.43 | 0.118 |

Table 1: Average (over the entire forecast interval) L2 loss values for the FC and the LSTM ResNet multiscale neural networks trained using different recursive step-ahead-training horizons ($n_f$) for the Van der Pol oscillator.

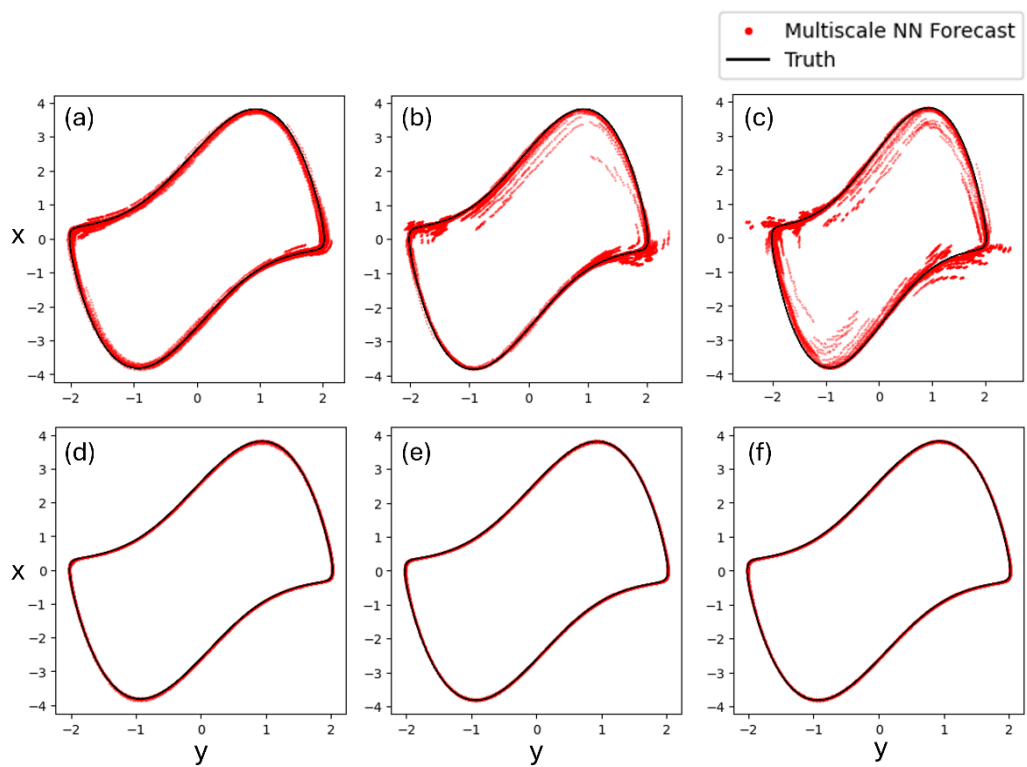

Figure 3: Phase-space ($x - y$) representation of forecasted trajectories from multi-step neural networks trained using various step-ahead horizons compared against the ground-truth trajectory for the Van der Pol oscillator: (a), (b), and (c) correspond to FC networks trained using 1-step, 4-step, and 8-step-ahead training horizons, respectively. (d), (e), and (f) show the corresponding results for the MLST-based architectures.

By extending the prediction window during training up to a certain point, this approach mitigates the risk of overfitting by enforcing the model to perform well across multiple time steps. This constraint encourages the model to learn more generalizable patterns and trends, increasing the potential for enhanced performance beyond the training window. However, if the prediction window is extended too far, the model may struggle with accumulating errors, making training less effective and potentially destabilizing the model's learning process. The results in Figure 2 (panel (b) vs (a) and panel (d) vs (c)) show that models trained using the "multi-step-ahead" training strategy generally achieve better performance and maintain stability over a longer duration compared to those trained using the often-adopted single-step-ahead approach. However, despite this improvement, cumulative prediction errors can still occur over longer horizons. While the "multi-step-ahead" training approach reduces the mismatch between training and forecasting, it does not entirely eliminate prediction error divergence.

The hierarchical multiscale method presents a more robust solution by leveraging multiple time scales to balance predictions' accuracy and stability. The last rows in each panel of Figure 2 demonstrate this method, where forecasts from different step-size models are combined to produce a stable long-term trajectory. This multiscale framework effectively reduces the number of recursive evaluations in time, thus mitigating the issue of accumulating errors.

A key advantage of the multiscale approach is its suitability for chaotic systems, where small errors can cause rapid deviation of the forecasted trajectory to deviate from reality. By incorporating various time-scale components, the hierarchical method ensures that errors do not accumulate as rapidly, preserving the longer-term trajectories on chaotic attractors.

To conclude, the LSTM-based architectures overall demonstrate smoother and more stable predictions compared to fully-connected ResNets. The values of the forecast losses from the multiscale models presented in Table 1 indicate that the LSTM-based architectures significantly outperform FC ResNets. The phase-space plots in Figure 3 further support this improvement. While multiscale models with standard FC ResNet layers remain stable, the trajectory deviates significantly from the true limit cycle. However, multiscale models with LSTM-based ResNet layers closely track the actual trajectory, demonstrating their effectiveness in preserving the system's long-term behavior.

Accordingly, throughout the rest of this paper, we use LSTM ResNets as the individual constituent neural network components of the hierarchical multiscale architecture.

**Section 4: Results**

We begin this section by presenting in subsection 4.1 the performance of the multiscale architecture on two additional canonical dynamical systems. We then proceed in subsection 4.2 to demonstrate the application of the multiscale architecture in two plasma test cases.

It is important to highlight that all the results provided in this paper are obtained using recursive (autonomous) mode for either generating the training sequence or forecasting the unseen test data by using the multiscale architectures' own previous predictions as inputs for future steps.

**4.1. Canonical Dynamical Systems**

To further evaluate the multiscale forecasting technique, we tested it on two additional canonical systems: the Cubic Oscillator and the Lorenz system. Together with the Van der Pol oscillator discussed in Section 3, these systems provide a set of representative benchmarks across different dynamical behaviors.

Unlike the Van der Pol oscillator, which exhibits periodic behavior with a closed limit cycle, the Cubic Oscillator features amplitude-dependent frequency, resulting in transient, state-dependent oscillations. This makes it a suitable test case for assessing the model's adaptability to non-periodic dynamics.

The Lorenz system is a classic example of chaotic dynamical behavior, characterized by its butterfly-shaped attractor and extreme sensitivity to initial conditions. The Lorenz system represent a particularly useful case for evaluating how long a forecast can remain accurate before diverging from the true trajectory in the presence of chaos. While exact trajectory predictions eventually deviate due to chaos, a well-trained model should ensure that the forecasted trajectory remains confined on the attractor in the phase space. This indicates that, even if precise state forecast is lost, the model successfully captures the system's overall dynamical structure, reconstructing the true attractor.

The details of data generation for all canonical systems are provided in Appendix A. A single trajectory comprising 1,000,000 data points was used for training, while testing was conducted on a separate trajectory of the same number of data points with different initial conditions. We trained three-layer LSTM-based ResNets, with various step sizes of $TS = 1, 2, 4, \ldots, 2024$. Each LSTM layer has 256 hidden states as stated in Appendix A. The multiscale architecture is then formed by selecting a subset of individual sub-networks (sub-models) trained using different step sizes.

### 4.1.1. Cubic Oscillator

After training sub-models independently with different step sizes, we can aggregate various sets of these individual networks to form a multiscale architecture. This aggregation allows the model to leverage both short-step sub-models for fine-scale accuracy and long-step sub-models for extended forecasting capability.

The forecasts of multiscale models with different set of sub-models are presented in Figure 4 and Figure 5, as well as in Table 2.

We highlight that in Figure 4, only the first 500,000 time steps are shown, whereas Figure 5 and the loss calculations in Table 2 consider the entire test sequence. The results are provided for two different step-ahead training strategies (single-step and 4-step represented by $n_f = 1$ and $n_f = 4$, respectively) for individual sub-models. Table 2 also present the corresponding losses for a more extreme case of $n_f = 8$.

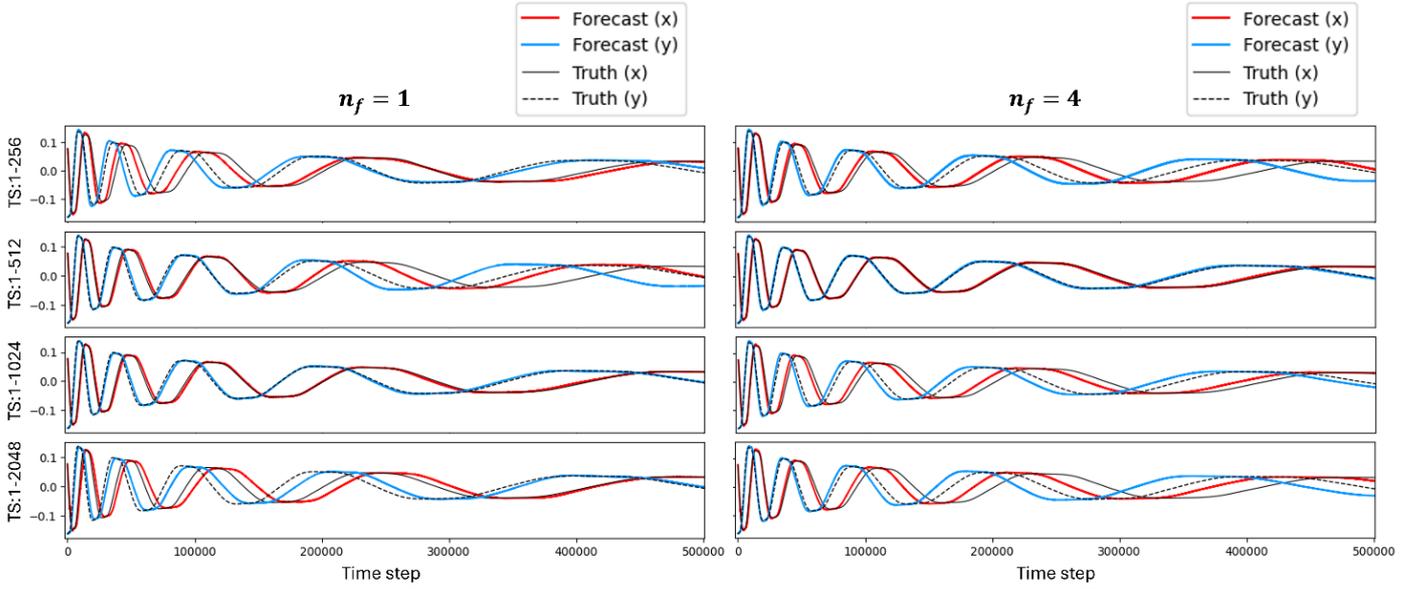

Figure 4: Forecasts from the multiscale network trained using 1-step-ahead (**left-column**) and 4-step-ahead (**right-column**) training strategy for the Cubic Oscillator case. Each row represents a multiscale architecture with different number of sub-models, comprised by models with step sizes $TS = 1, 2, …, TS_{max}$, where $TS_{max}$ denotes the sub-model with the maximum step size for each row.

A key observation from the presented results is the impact that the number of sub-models and the maximum step size used for training have on forecasting accuracy. It is indeed noticed that inclusion of larger step sizes beyond a certain point during training can introduce error in the predictions.

In this respect, the multiscale architecture trained using a 4-step-ahead strategy, with sub-models having step sizes up to $TS = 512$, achieves the best performance, yielding an average loss of 0.35 % as reported in Table 2.

For single-step-ahead training, the architecture with sub-models extending up to $TS = 1024$ performs best, resulting in a loss of 4.86 %. This behavior likely occurs because, at very large step sizes, the network operates farther from the governing physical equations, and the sampling of dynamics becomes too sparse relative to the system's characteristic time scales, making it harder to learn a generalizable function. This suggests that while using more sub-models with larger step sizes reduces the number of recursive evaluations over time (and possibly error accumulation during recursive evaluation process), there remains a trade-off between step size and forecasting accuracy.

Moreover, it is observed that models trained using different recursive training strategies ($n_f = 1, 4, 8$) exhibit varying levels of predictive accuracy, demonstrating the impact of the selection of the training horizon.

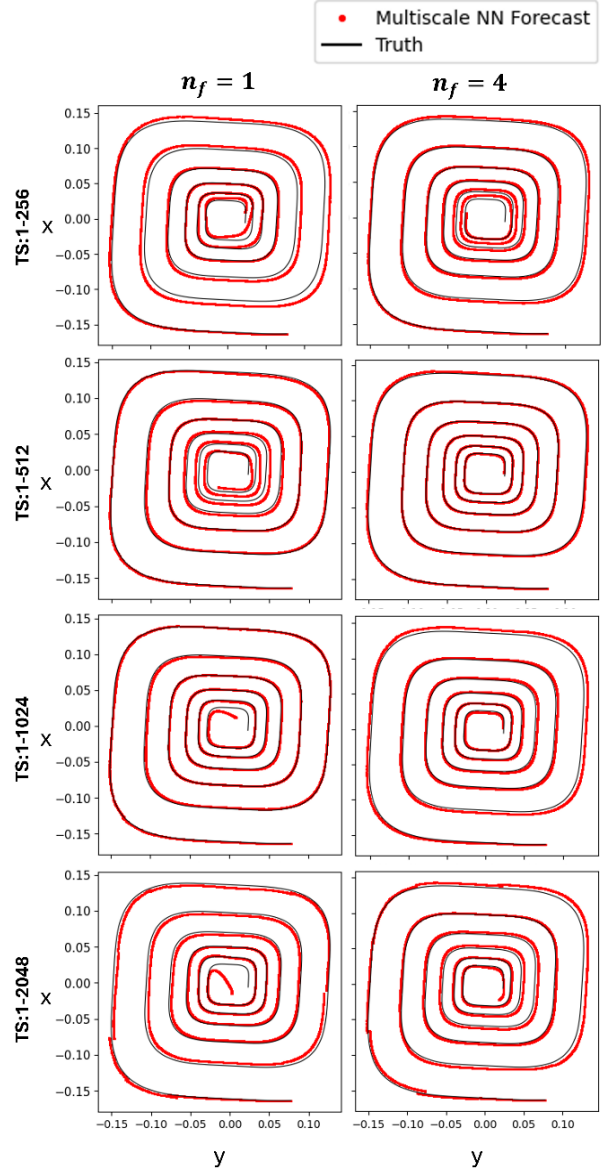

Figure 5: Phase-space ($x - y$) representation of the forecasted trajectories from multistep neural networks trained using 1-step-ahead (**left column**) and 4-step-ahead (**right column**) training strategy for the Cubic Oscillator problem. Each row represents a multiscale architecture consisting of different number of sub-models with step sizes TS = 1, 2, …, $TS_{max}$, where $TS_{max}$ denotes the sub-model with the maximum step size included.

|  | Loss [%] | | |
| --- | --- | --- | --- |
|  | $n_f = 1$ | $n_f = 4$ | $n_f = 8$ |
| **TS:1-256** | 20.881 | 74.003 | 13.702 |
| **TS:1-512** | 62.308 | 0.353 | 0.799 |
| **TS:1-1024** | 4.857 | 10.939 | 57.204 |
| **TS:1-2048** | 20.814 | 34.106 | 3.650 |

Table 2: Average (over the entire forecast interval) L2 loss values for multiscale neural networks trained using different recursive step-ahead training horizons ($n_f$) and including different subset of sub-models for the Cubic Oscillator problem. The losses are provided for multiscale architectures consisting of different number of sub-models with step sizes TS = 1, 2, …, $TS_{max}$, where $TS_{max}$ denotes the sub-model with the maximum step size included.

Additional insights are provided by Figure 6, which presents the forecasts generated by individual sub-models with different step sizes, highlighting their inability to maintain consistency with the ground-truth over extended forecasting horizons.

The forecasts generated by these single step-size models initially align well with the true trajectory but gradually diverge over time. This deviation is not only observed when forecasting an unseen trajectory but also when generating the training data itself as shown in Figure 7, where the model recursively solves future steps using its own outputs as inputs. This emphasizes the fundamental limitations of single step-size networks for long-term predictions.

A general observation from these two figures is that networks trained with smaller step sizes exhibit a faster divergence, as seen in the top rows of both Figure 6 and Figure 7, while networks trained with larger step sizes tend to deviate more slowly. This aligns with the expectation that larger step sizes result in less frequent recursive evaluations, meaning that smaller errors accumulate over time.

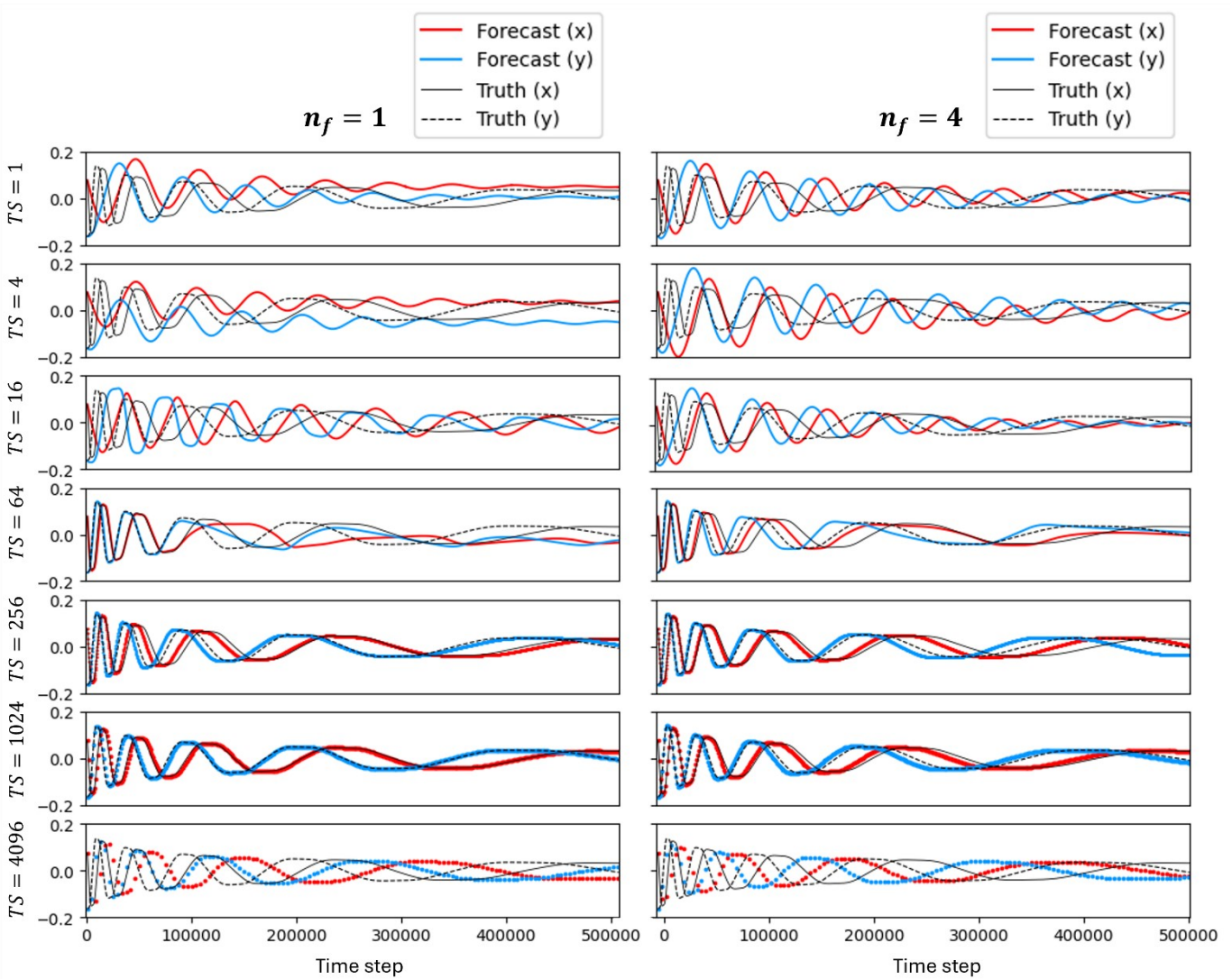

Figure 6: **Forecasts** from individual sub-models with different step sizes compared against ground-truth trajectories for the Cubic Oscillator problem: (**left column**) 1-step-ahead and (**right-column**) 4-step-ahead training strategy. Each row represents predictions from a single-time-step model with the step size $TS$.

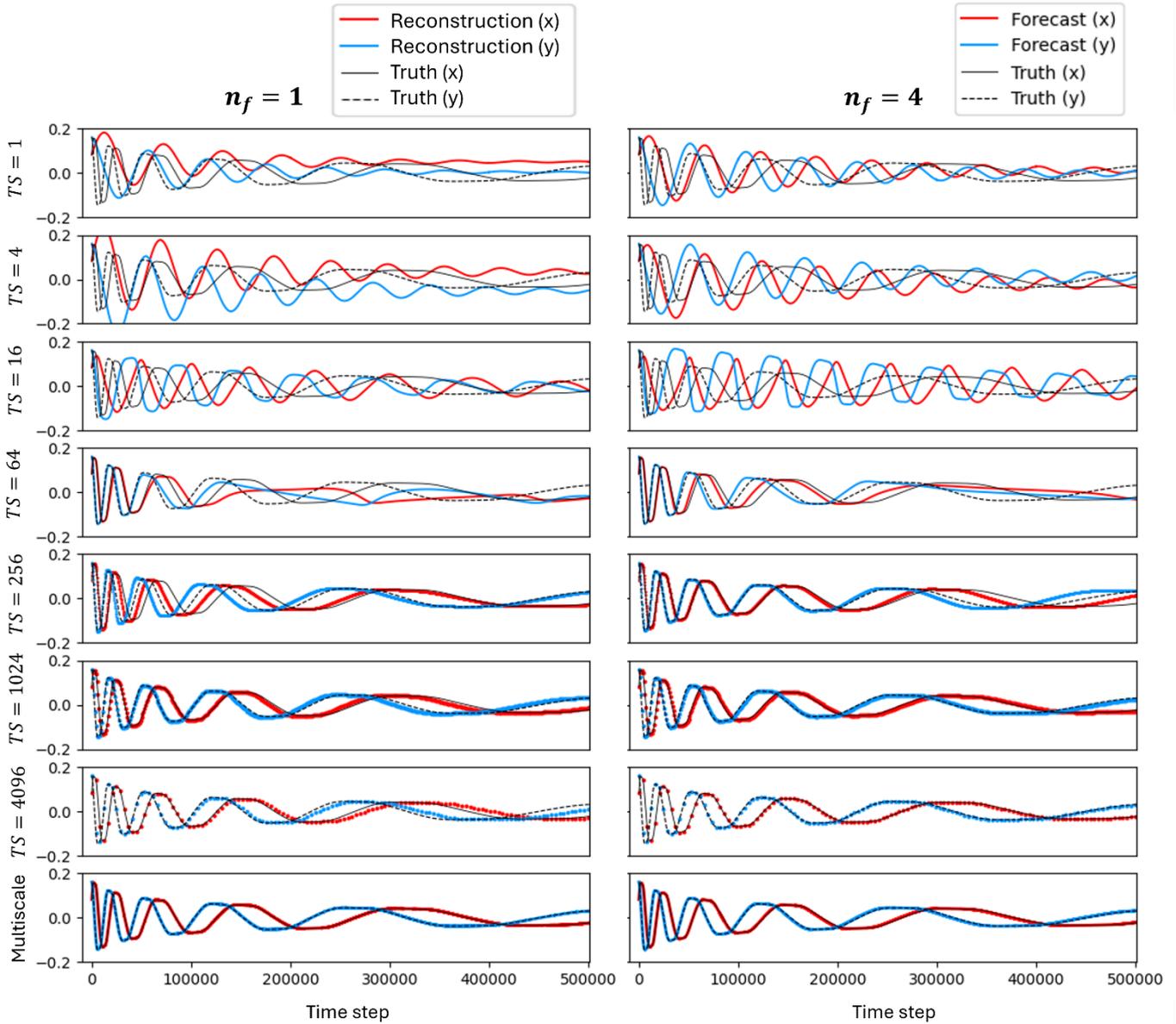

Figure 7: **Reconstruction** of the training data sequence from individual sub-models with different step sizes (TS) compared against ground-truth trajectories for the Cubic Oscillator problem: (**left column**) 1-step-ahead and (**right-column**) 4-step-ahead training strategy. Each row represents predictions from a single-time-step model with the step size $TS$. The bottom-most row represents the reconstruction from the multiscale architecture which incorporates sub-models with step size $TS = 1 – 4096$.

### 4.1.2. Lorenz System

Being a chaotic dynamical system, the Lorenz problem poses additional challenges in long-term forecasting. The analysis here explores the forecasting performance of the hierarchical multiscale architecture in the presence of chaos. The sub-model networks in this section are trained using a *2-step-ahead* training strategy.

Figure 8 shows that the forecast from the multiscale architecture, which integrates sub-models trained with step sizes up to $TS = 128$, closely follows the ground truth for an extended period. The predictions of the three state variables – $x, y$ and $z$ – initially align well with the true trajectory but begin to deviate after approximately 7,500 timesteps.

Despite the trajectory-level discrepancies, Figure 9 demonstrates that the forecasts remain on the attractor, indicating that the model successfully captures the system's underlying nonlinear dynamics. While minor distortions are present, the multiscale approach effectively preserves the global structure of the Lorenz system, even if maintaining exact trajectory accuracy over long time-horizons remains challenging.

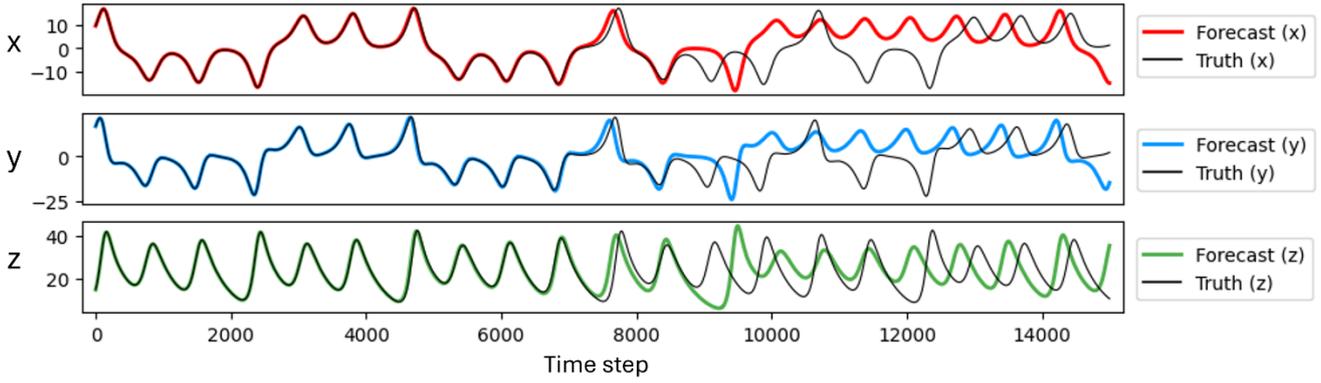

Figure 8: Forecasts from the multiscale architecture consisting of models with step sizes TS = 1, 2, …, 128 and trained using 2-step-ahead training strategy for the Lorenz system. Each row represents a state variable of the system $x$, $y$ and $z$.

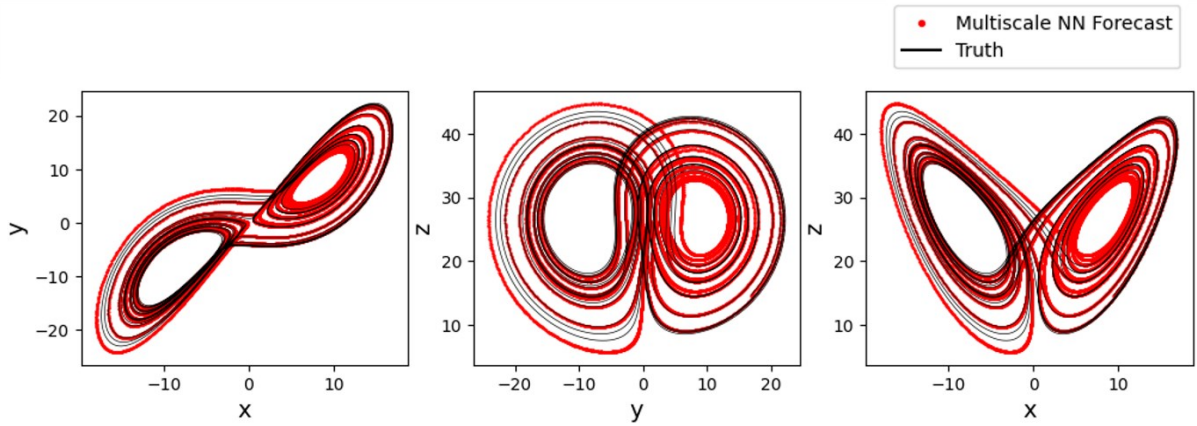

Figure 9: Phase-space ($x - y - z$) representation of the forecasted trajectories from the multiscale architecture consisting of models with step sizes TS = 1, 2, …, 128 and trained using 2-step-ahead training strategy for the Lorenz system.

The individual sub-model forecasts for different step sizes are provided in Figure 10. As observed in the previous test cases, sub-models with smaller step sizes suffer from gradual error accumulation, due to recursive feedback, and diverge early. In contrast, temporally sparse forecasts of sub-models with moderate step sizes remain in line with the true trajectory for longer times, which are covered with fewer recursive evaluations. However, sub-models with excessively large step sizes ($TS \geq 1024$) again exhibit early divergence and performance degradation, emphasizing the challenge of capturing generalized long-term behavior when data points are too sparsely distributed.

Figure 11 and Figure 12 illustrate the impact of step-size selection and sub-model aggregation on forecasting accuracy for the Lorenz system, where the forecasts of different multiscale architectures incorporating sub-models with varying maximum step sizes ($TS_{max}$) are presented.

The results demonstrate that a direct correlation exists up to a certain point between how much the forecasting horizon can be extended and what maximum step size can be used for the constituent sub-models of a multiscale architecture.

The multiscale model with $TS_{max} = 128$ shows the most extended prediction horizon. Increasing $TS_{max}$ beyond this value limits the prediction horizon as the accuracy of the sub-models with larger step sizes deteriorates. It is also observed that incorporating excessively large step-size sub-models into the multiscale architecture distorts the trajectory, resulting in non-smooth predictions and, in some cases, causing the forecasted trajectory to even deviate from the attractor.

These observations highlight that selecting an appropriate step-size range is crucial for effectively capturing the dynamics (learning the attractor) of chaotic systems, like the Lorenz.

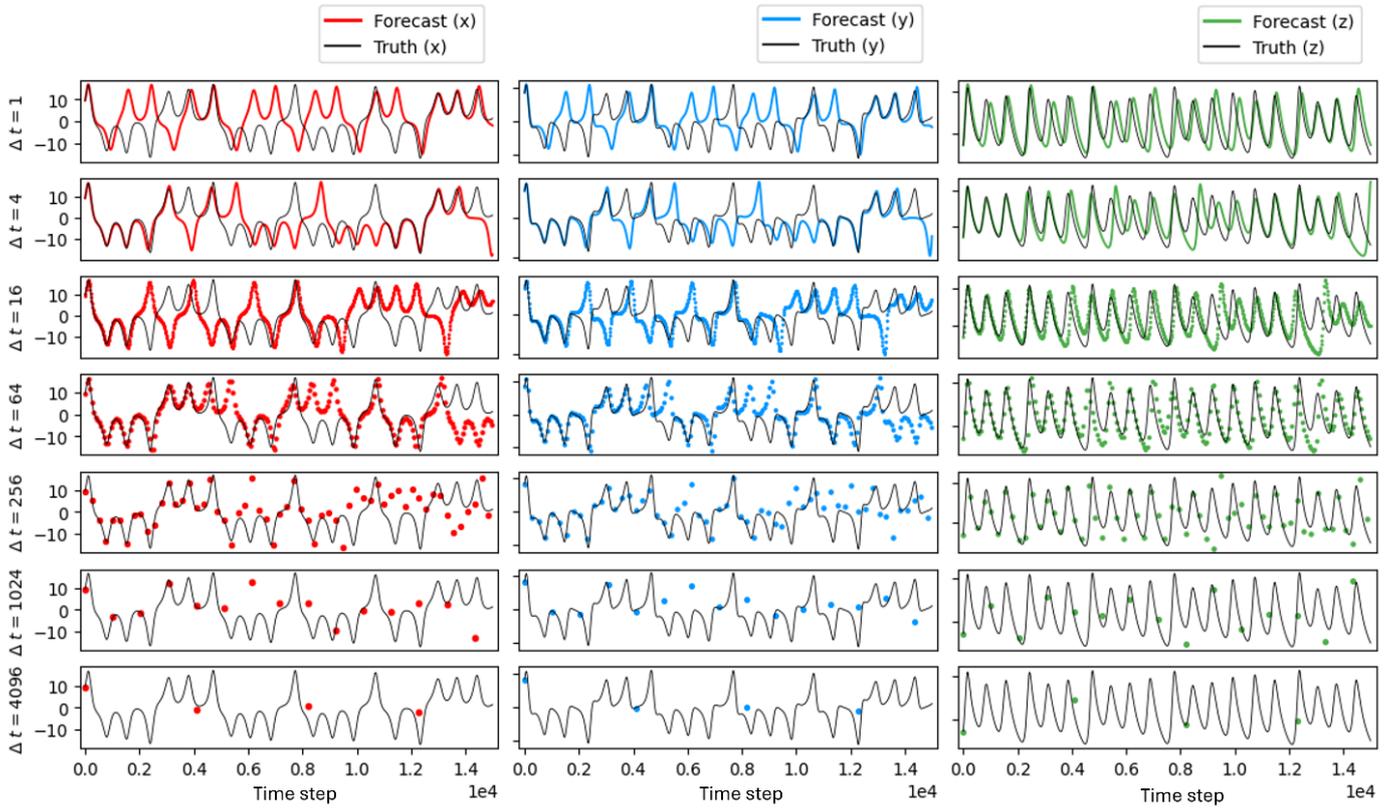

Figure 10: Forecasts from individual sub-models with different step sizes compared against the ground-truth trajectories for the Lorenz system. Each column represents a state variable of the system $x$, $y$ and $z$.

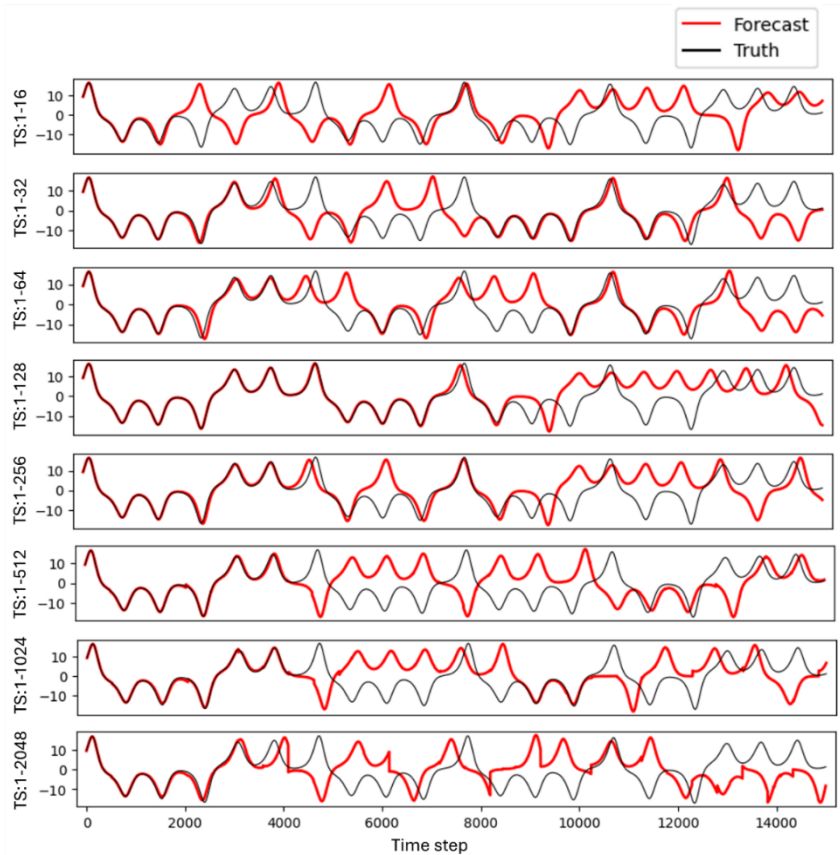

Figure 11: Forecasts from various multiscale architectures for the Lorenz system. Each row represents a multiscale architecture with different number of sub-models, including models with step sizes TS = 1, 2, …, $TS_{max}$, where $TS_{max}$ denotes the sub-model with the maximum step size on that row.

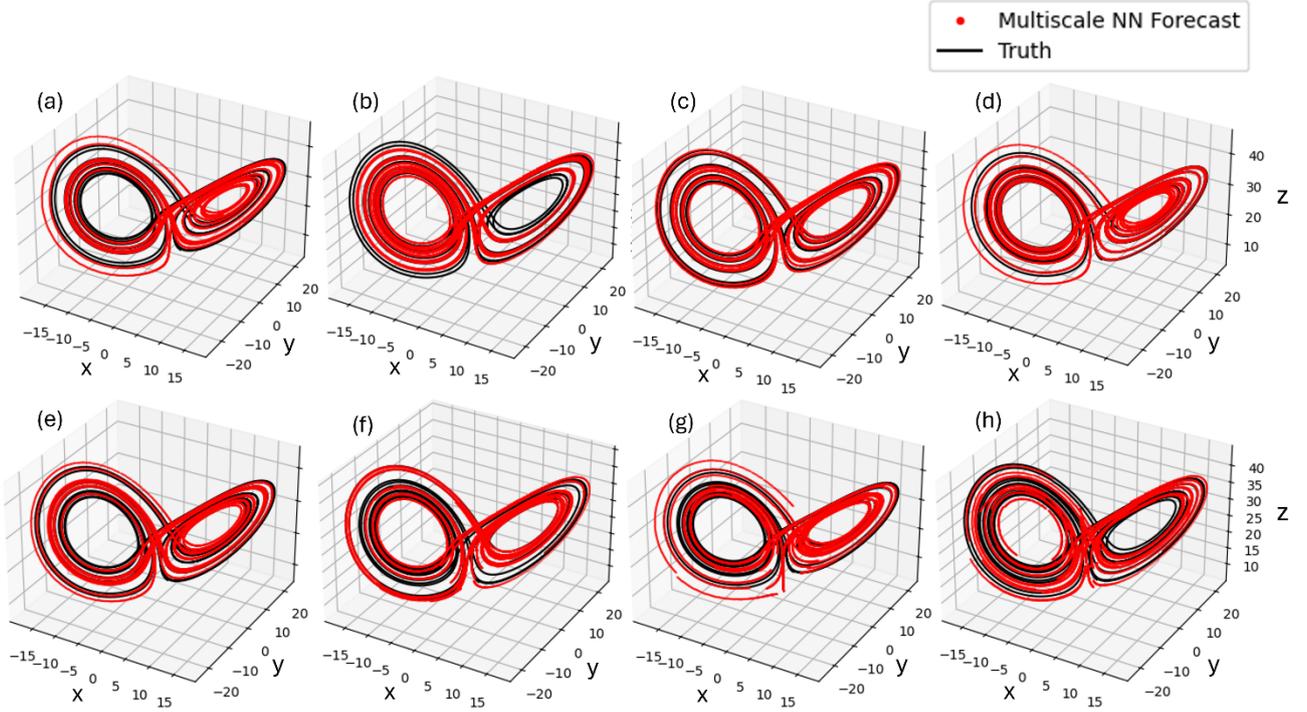

Figure 12: Phase-space ($x - y - z$) representation of the forecasted trajectories from various hierarchical multiscale networks consisting of different number of sub-models with step sizes TS = 1, 2, ..., $TS_{max}$, where $TS_{max}$ denotes the sub-model with the maximum step size. (a) $TS_{max} = 16$, (b) $TS_{max} = 32$, (c) $TS_{max} = 64$, (d) $TS_{max} = 128$, (e) $TS_{max} = 256$, (f) $TS_{max} = 512$, (g) $TS_{max} = 1024$, and (h) $TS_{max} = 2048$.

## 4.2. Plasma test cases

Following all the above demonstrations and discussions, we now apply the multiscale architecture to plasma test cases in order to assess how the approach addresses challenges such as error accumulation, scale-dependent accuracy, and the preservation of dynamical features across different scales for complex plasma systems. We focus on two E × B (cross-field) plasma configurations, which are characterized by the presence of mutually perpendicular electric ($E$) and magnetic ($B$) fields. These plasmas are common across various industrially important applications, including Hall thrusters for spacecraft propulsion and magnetrons for material processing.

The neural network models are trained on plasma data obtained from kinetic particle-in-cell simulations [48][49] and are employed to simultaneously predict various two-dimensional (2D) distribution of plasma properties of interest specified for each case. Data are generated using the IPPL-Q2D PIC simulation code [50], which is based on the reduced-order PIC scheme developed by the authors [51]. The code is extensively benchmarked across different test cases [50]-[53].

Given the high dimensionality of the data, we use a lower-rank approximation instead of raw data, preserving essential information while reducing complexity. This is achieved by applying Singular Value Decomposition (SVD) [54] to the dataset for each plasma property, keeping only the first $r$ singular values and their corresponding modes. The appropriate value of $r$ is case-specific which depends on the singular values distribution, balancing sufficient information retention with the desired level of dimensionality reduction.

Consider $X_p \in R^{n \times m}$ to be a matrix containing the spatiotemporal data of a single plasma property $p$ such that each column represents flattened spatial distribution (snapshot) of the plasma property at a certain time. Hence, $n$ is the spatial dimension and $m$ is the number of time instants. The SVD of matrix $X_p$ is given by

$$X_p = U_p \Sigma_p V_p^T. \tag{Eq. 1}$$

The columns of matrix $U$ represents the SVD spatial modes, forming an orthonormal basis for representing the data matrix $X_p$. Columns of $V$ represent the amplitudes of these modes over time, thus capturing temporal patterns. $\Sigma$ is a diagonal matrix with non-negative singular values arranged in descending order, which indicate the importance or strength of the corresponding singular vectors in $U$ and $V$. Accordingly, the rank-$r$ approximation ($\tilde{X}_p^r$) of the data is obtained by retaining the first $r$ columns of $U$ and $V$ and the first $r$ singular values in the SVD expansion

$$\tilde{X}_p^r = \sum_{i=1}^{r} u_i \sigma_i v_i^T = \tilde{U}_p^r \tilde{\Sigma}_p^r \tilde{V}_p^{rT}. \tag{Eq. 2}$$

$\tilde{V}_p^{rT} \in R^{m \times r}$ represent the reduced-dimension (rank-$r$ truncated) representation of the plasma property $p$ in the SVD basis. The complete state vector of the system ($\tilde{V}^{rT} \in R^{m \times r \times n_p}$, where $n_p$ is the number of plasma properties), aggregates the reduced-dimension approximations of all relevant plasma properties which is expressed as

$$\tilde{V}^{rT} = \left[\tilde{V}_1^{rT}, \tilde{V}_2^{rT}, \dots, \tilde{V}_{n_p}^{rT}\right]. \tag{Eq. 3}$$

The truncated SVD modes ($\tilde{U}_p^r$) and the singular values ($\tilde{\Sigma}_p^r$) for each property are derived using training data. During the forecast phase, we predict the reduced-dimensional state vector ($\tilde{V}^{rT}$) and then reconstruct the spatial distribution of each plasma property using the respective SVD modes and singular values according to Eq. 2.

### 4.2.1. 2D radial-azimuthal $E \times B$ plasma configuration

This test case resembles a radial-azimuthal cross-section of an $E \times B$ discharge configuration, closely following the simulation case descriptions in Ref. [55]. The computational domain is a 2D Cartesian plane with dimensions of 1.28 cm along both simulation directions ($L_x = L_z = 1.28\ cm$), where the $x$, $y$, and $z$ axes correspond to the axial, radial, and azimuthal directions, respectively. The computational mesh consists of cells with a size of 50 $\mu m$, yielding 256 nodes along each simulation direction. A constant axial electric field ($E_x$) of 10 $kVm^{-1}$ and an external radial magnetic field ($B_y$) of 20 mT are imposed.

Initially, electrons and ions are loaded with a uniform distribution across the domain according to Maxwellian distributions with temperatures of 10 eV and 0.5 eV, respectively. The initial plasma density is set at $1.5 \times 10^{16}\ m^{-3}$, with each computational cell containing 100 simulation macroparticles. Collisions are neglected and sustainment of a steady-state discharge is achieved through a particle injection source. The injection source is azimuthally uniform but follows a cosine profile radially, extending from $y = 0.09\ cm$ to $y = 1.19\ cm$, with a peak of $8.9 \times 10^{22} m^{-3} s^{-1}$. Electron-ion pairs are sampled from Maxwellian distributions corresponding to the species' initial temperatures and are injected according to the specified source profile.

Boundary conditions for particles are set as follows: particles reaching the radial walls are removed, with no secondary electron emission considered. To enforce azimuthal periodicity, particles crossing the azimuthal boundaries are reintroduced at the opposite side with unchanged velocity and radial coordinates. Since the axial direction is not resolved, an artificial axial extent of 1 cm is assumed on both sides of the simulation plane [50][55]. Particles reaching the axial boundaries are re-sampled from the initial Maxwellian distributions and reloaded at the same radial and azimuthal coordinates.

For the electric potential, a zero-volt Dirichlet boundary condition represents the grounded radial walls, while a periodic condition is imposed along the azimuthal boundaries.

The simulation is conducted using the IPPL-Q2D code with a domain decomposition of 50 regions along both the radial and azimuthal directions in order to enable the reduced-order problem treatment [51]. This level of approximation has been demonstrated to produce results consistent with full 2D simulations while yielding about 5 times reduction in computational cost [50].

A time step of $1.5 \times 10^{-11}$s is used, and simulation outputs are averaged over 1,000 timesteps for a total simulated time of about 135 $\mu s$, yielding a dataset of about 9,000 snapshots in time for each plasma property. The initial 350 snapshots representing the system's transient behavior are discarded. The following 6,000 frames are used for training, and the remaining data is kept for testing (about 2,650 snapshots). The plasma properties of interest are the 2D distributions of the electron number density ($n_e$), the electrons' axial, radial, and azimuthal velocities ($v_{ex}$, $v_{ey}$ and $v_{ez}$, respectively), the azimuthal electric field ($E_z$), and radial and azimuthal electron temperatures ($T_{ey}$ and $T_{ez}$, respectively).

Figure 13 presents the distribution of singular values for each plasma property in this test case as derived from the training dataset. Based on the observed distributions, the first 30 SVD modes were retained to approximate the data.

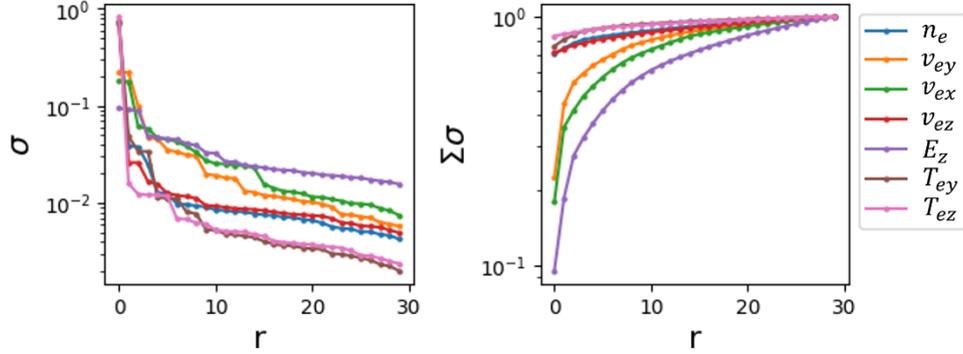

Figure 13: Distributions of (**left**) normalized singular values ($\sigma$) and (**right**) normalized cumulative sum of the first r dominant singular values ($\Sigma\sigma$) from the SVD of various plasma properties for the radial-azimuthal E × B plasma test case.

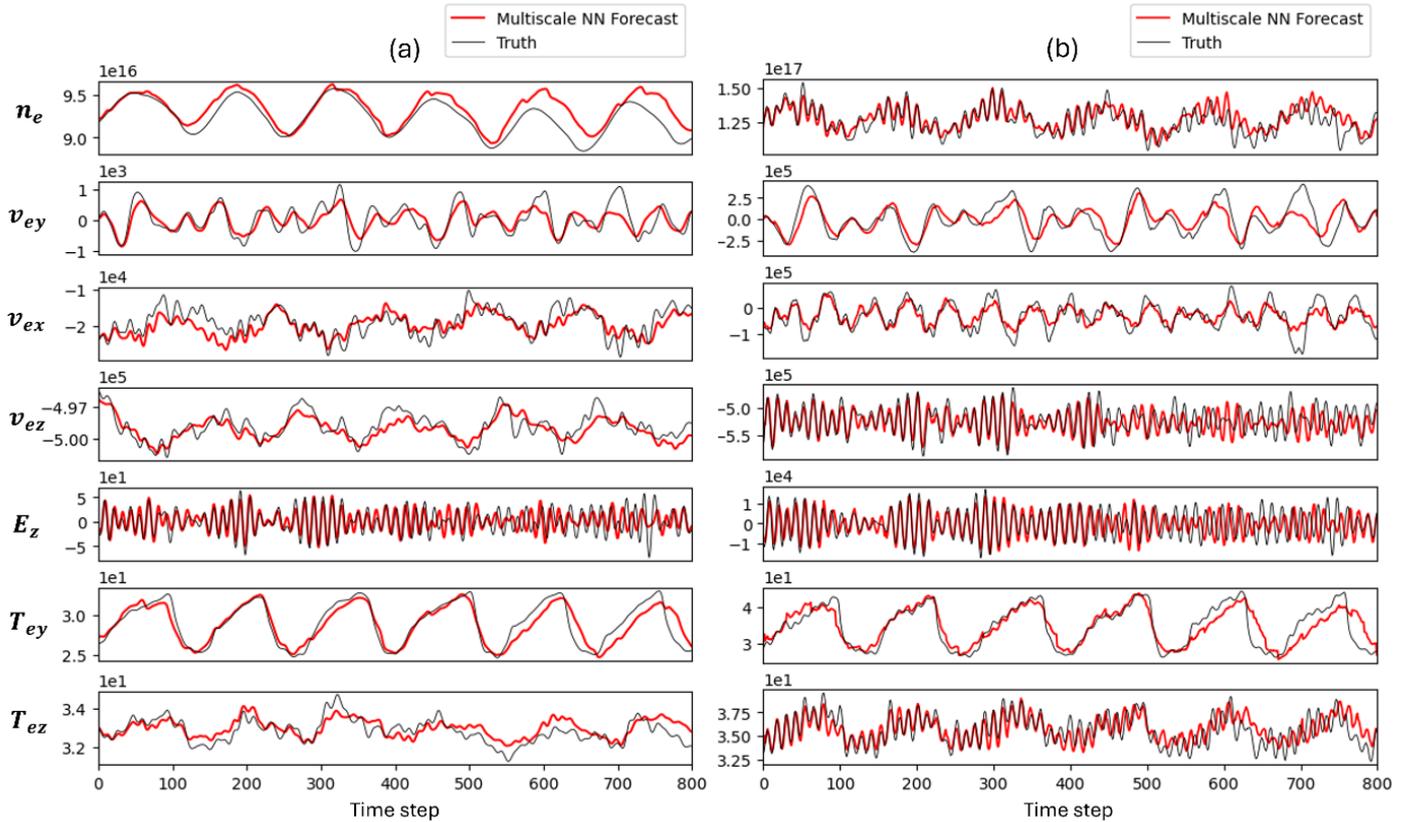

Figure 14: Forecasts from the multiscale network consisting of sub-models with step sizes $TS = 1 - 64$ for the radial-azimuthal E × B plasma test case: time evolutions of (**a**) spatially averaged and (**b**) mid-domain local values of different plasma properties, namely, electron number density ($n_e$), electrons' axial, radial, and azimuthal velocities ($v_{ex}$, $v_{ey}$ and $v_{ez}$, respectively), azimuthal electric field ($E_z$), and radial and azimuthal electron temperatures ($T_{ey}$ and $T_{ez}$, respectively).

For the present test case, we used three-layer LSTM-based ResNets with various step sizes of $TS = 1, 2, 4, \ldots, 256$. Each LSTM layer consists of 800 hidden states, and training was conducted using a 5-step-ahead strategy. The multiscale architecture was then constructed by integrating a selected set of trained sub-models with different step sizes.

The forecasts of the resulting multiscale network comprising sub-models with step sizes from $TS = 1$ to $TS = 64$ are presented in Figure 14 to Figure 16. These figures show simultaneous predictions of the multiscale network for several plasma properties during the test interval. Figure 14 presents the time evolution of both spatial-average values and mid-domain local values. Figure 15 provides signal traces of the local plasma property values at two randomly chosen locations as additional information for comparisons between the forecast and the ground-truth data. Figure 16 illustrates sample predicted snapshots.

These figures indicate that the forecasts from the multiscale model generally follow the trends of the ground-truth data from the PIC simulation, although some deviations become more noticeable toward the end of the displayed interval. Moreover, the snapshots clearly show that the dominant spatial structures in plasma properties are preserved, demonstrating the multiscale architecture's forecasting capability not only in the temporal domain but also toward capturing the spatial complexities inherent in the radial-azimuthal $E \times B$ configuration of the present test case.

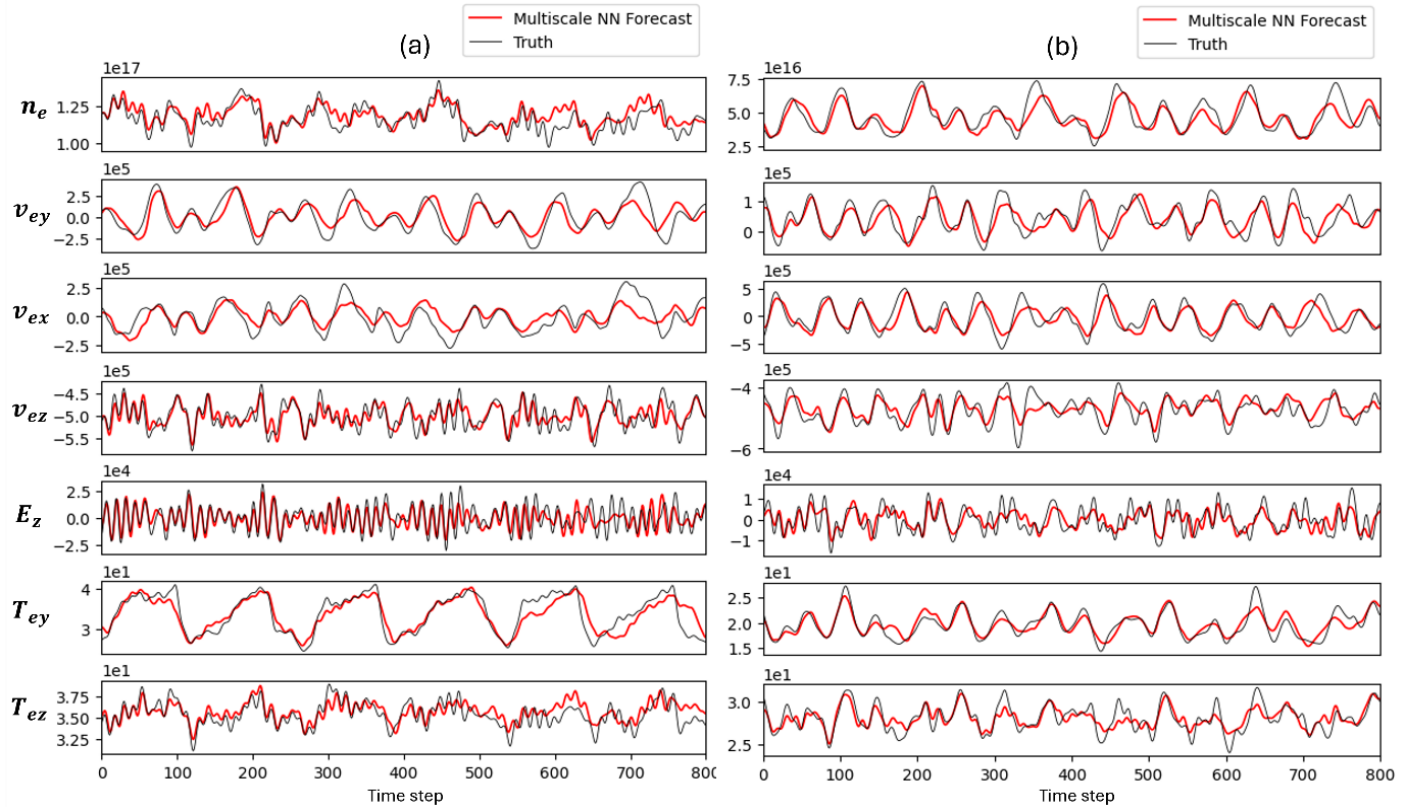

Figure 15: Forecasts from the multiscale network consisting of models with step sizes $TS = 1 - 64$ for the radial-azimuthal $E \times B$ plasma test case: time evolutions of the local values of different plasma properties at two random locations in the domain represented by columns (a) and (b). The forecasted properties include electron number density ($n_e$), electrons' axial, radial, and azimuthal velocities ($v_{ex}$, $v_{ey}$ and $v_{ez}$ respectively), azimuthal electric field ($E_z$), and radial and azimuthal electron temperatures ($T_{ey}$ and $T_{ez}$, respectively).

Figure 17 provides the reconstructions of the training data and the forecasts over the test interval using individual sub-models with varying step sizes ($TS$) for the azimuthal electric field ($E_z$) as a representative property. Similar results for other plasma properties are provided in Appendix B. Comparing the forecasts from individual sub-models to the aggregated multiscale forecasts, we observe a pattern similar to that seen for the canonical systems. In particular, models with small TS (e.g., $TS \leq 4$) tend to accumulate errors quickly due to the high number of recursive evaluations. Their forecasts diverge from the ground-truth relatively early. In contrast, models with larger $TS$ show slower error accumulation because they require fewer recursions. However, these larger steps do not capture the finer temporal dynamics. Figure 17(a) indicates that that error accumulation in small step-size models occurs even during the reconstruction of the training dataset.

Nevertheless, the hierarchical multiscale approach effectively balances short-term accuracy with long-term stability. This is in line with our earlier discussions: by reducing the total number of recursive steps and leveraging complementary strengths of sub-models across scales, the multiscale model better approximates the true plasma dynamics. The reconstruction of the training data using the multiscale model in recursive mode for all plasma properties are provided in Appendix B. Ensuring that the model is able to generate the training sequence serves as a pre-assessment before it is challenged with forecasting.

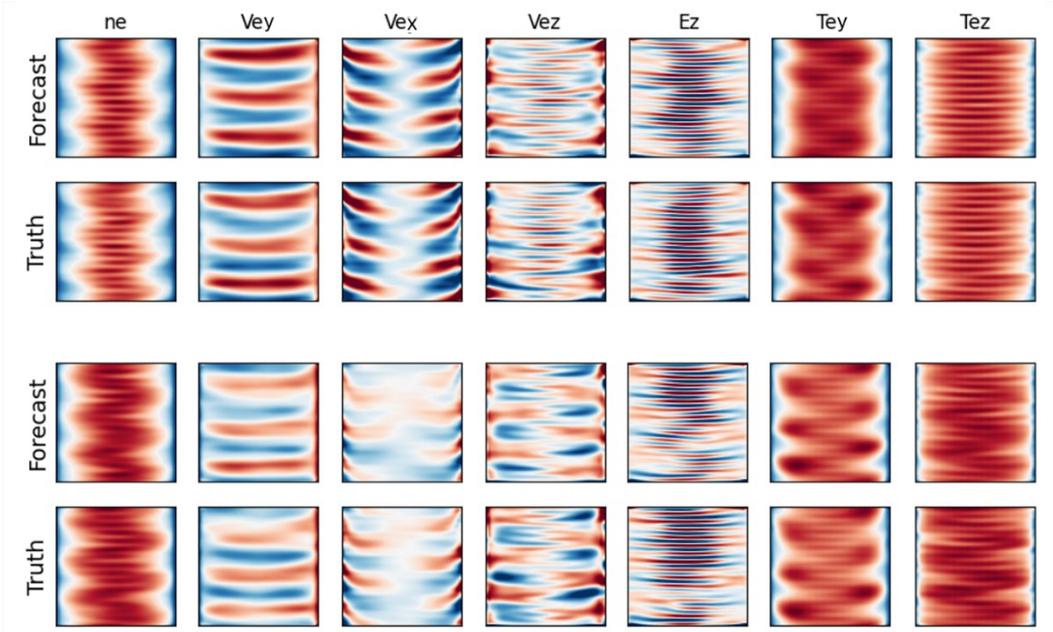

Figure 16: Forecasted 2D snapshots of different plasma properties from the multiscale network at two sample time instants within the test interval compared against the corresponding ground-truth snapshots from the PIC simulation. In all plots, the horizontal axis represents the radial coordinate ($y$), and the vertical axis represents the azimuthal coordinate ($z$). The columns from left to right correspond to electron number density ($n_e$), electrons' axial, radial, and azimuthal velocities ($v_{ex}$, $v_{ey}$ and $v_{ez}$ respectively), azimuthal electric field ($E_z$), and radial and azimuthal electron temperatures ($T_{ey}$ and $T_{ez}$, respectively).

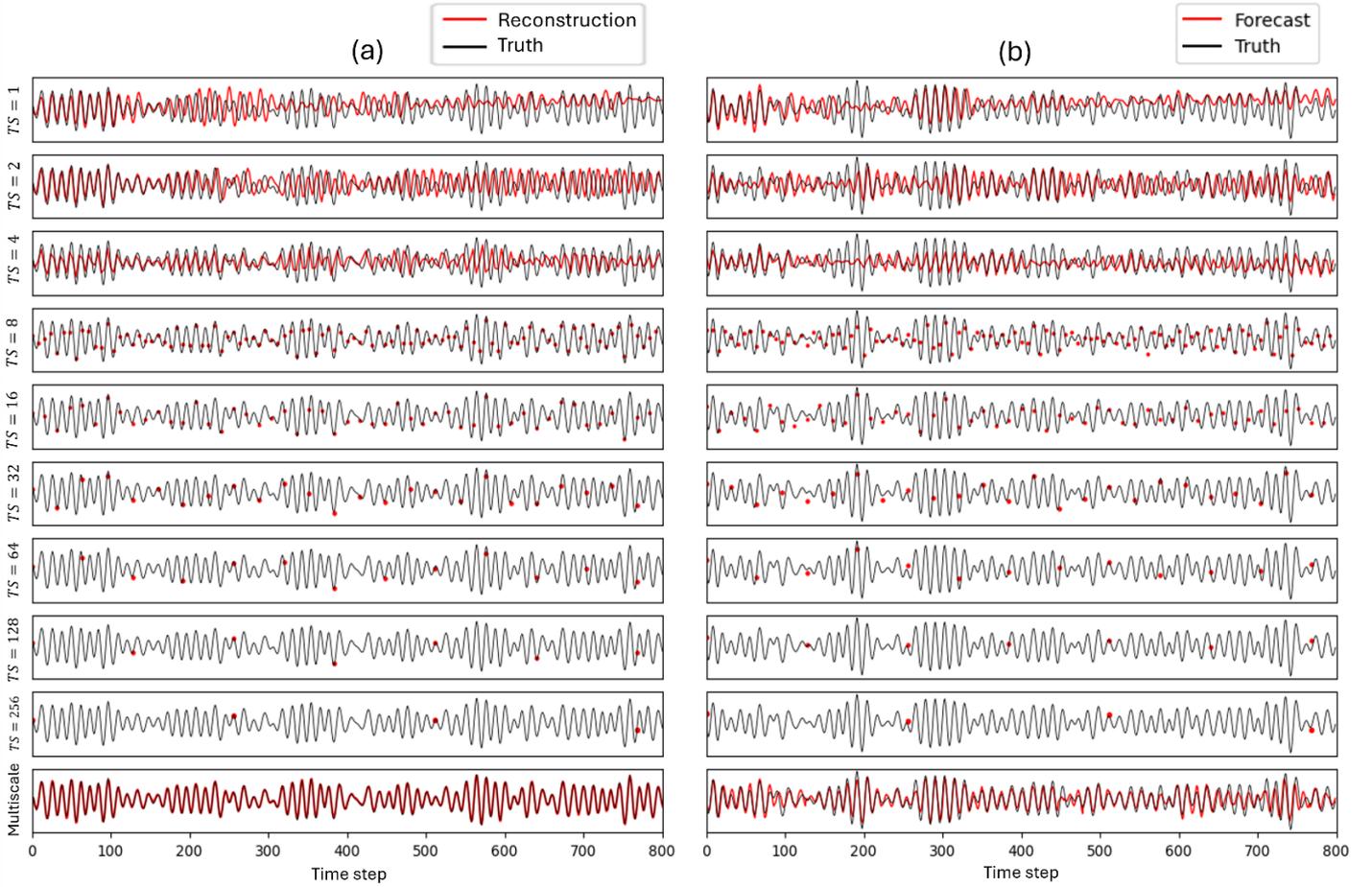

Figure 17: Time evolutions of the spatially averaged azimuthal electric field ($E_z$) for the radial-azimuthal E × B plasma test case: (**a**) reconstructions of the training data (**b**) forecasts of the testing data from individual sub-models with different step sizes and from multiscale model (bottom-most row) consisting of models with step sizes $TS = 1 - 256$ compared against the ground-truth data.

## 4.2.2. 2D axial-azimuthal E × B plasma configuration

The configuration of this test case represents a 2D Cartesian plane of a generic Hall thruster geometry along its axial-azimuthal coordinates. The setup of the problem is adopted from the benchmarking effort by Charoy et al. [56]. The domain's axial extent ($L_x$) is 2.5 cm, and the azimuthal length of the domain ($L_z$) is 1.28 cm. Consistent with the benchmark settings [56], the sizes of the computational cells for the ground-truth PIC simulation along both the axial and azimuthal directions are $\Delta x = \Delta z = 50\ \mu m$. This corresponds to the number of nodes of 500 along the axial ($x$) axis and 256 along the azimuthal ($z$) axis.

The collisional processes are not accounted for in this test case. To maintain a quasi-steady state in the system, a temporally constant ionization source is imposed according to the benchmark's definition in Ref. [56]. The axial distribution of the ionization source is a cosine function spanning over $x = 0.25$ cm to $x = 1$ cm, with the peak value of $6.54 \times 10^{22}\ m^{-3}s^{-1}$, which establishes an average ion current density of $J_M = 50\ Am^{-2}$. The applied magnetic field ($B$) is purely radial and has a Gaussian profile along the axial direction, similar to that in Ref. [56], with the peak intensity of 10 mT.

The discharge voltage applied between the anode and the cathode boundaries of the domain is 200V. Therefore, as the boundary condition of the potential solver, the 200-V and 0-V Dirichlet conditions are applied at the anode and the cathode boundaries, respectively.

Initialization of the simulations is performed by loading electrons and ions uniformly throughout the simulation domain as electron-ion pairs with the initial density of $5 \times 10^{16}\ m^{-3}$. The particles are randomly sampled from Maxwellian distribution functions at temperatures of 10 eV for the electrons and 0.5 eV for the ions. The initial macroparticle-per-cell count for each species is 100.

Regarding the particles' boundary conditions, both ions and electrons crossing either the anode or the cathode boundaries are removed from the simulation. To maintain the discharge, at each timestep, electrons are injected into the domain from the cathode side, with the number of injected electrons obtained using the quasi-neutrality condition at the cathode plane as described in Refs. [52][57]. The re-injected electrons are sampled from a Maxwellian distribution at 10 eV. Along the azimuthal direction, the periodic boundary condition is implemented such that the particles leaving the domain at one end are injected back from the other end while keeping their axial position and their velocity.

The simulation is performed using the IPPL-Q2D code, employing a domain decomposition associated with the reduced-order PIC using 20 horizontal regions ($N$) along the $z$ direction and 40 vertical regions ($M$) along the $x$ direction. It is demonstrated in Refs. [51][52] that, at the selected numbers of regions, the predictions of the reduced-order PIC converges to the full-2D results while providing about 12 times computational speedup.

The timestep of the simulations is $5 \times 10^{-12}$ s, with a total simulation duration of 30 $\mu s$. Data is averaged every 1,000 timesteps, resulting in 6,000 snapshots. The first 700 snapshots, which capture the system's transient behavior, are discarded, while the next 3,300 snapshots are used for training. The remaining snapshots are reserved for testing.

In the case study here, we focused on developing a multiscale model to capture variations in the azimuthal electric field ($E_z$). Therefore, only the snapshots of the azimuthal electric field constitute the dataset. The data matrix is approximated using the 50 leading SVD modes and their corresponding singular values, as shown in Figure 18.

Furthermore, we utilized three-layer LSTM-based ResNets with step sizes of $TS = 1, 2, 4, \ldots, 256$ as sub-models of the multiscale architecture. Each LSTM layer comprised 128 hidden states, and training was performed using a 3-step-ahead prediction approach.

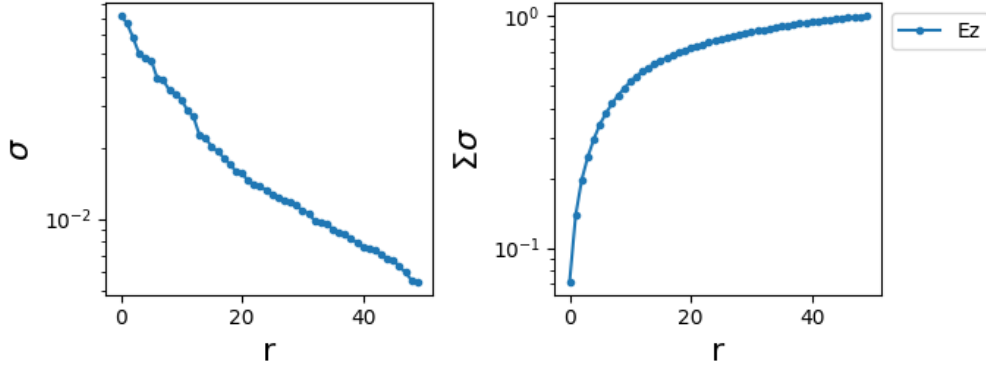

Figure 18: Distributions of (**left**) normalized singular values ($\sigma$) and (**right**) normalized cumulative sum of the first $r$ dominant singular values ($\Sigma\sigma$) from the SVD of azimuthal electric field for the axial-azimuthal E × B plasma test case.

The results, presented through Figure 19 to Figure 21, assess the forecasting performance of the multiscale neural-network model for the axial-azimuthal plasma test case.

In particular, Figure 19 provides the time evolution of the spatially averaged $E_z$, showing reconstructions in the training interval and forecasts in the test interval across several multiscale networks, each comprised by sub-models of different step sizes. These plots show that all different multiscale models generate training data with differences that are nearly indistinguishable from one another. During forecasting, their performance exhibits subtle variations but remains largely consistent both among each other and compared against the ground-truth data. Toward the end of the forecasted sequence, however, deviations become more pronounced, highlighting the inherent limitation on how far into the future reliable predictions can extend.

Figure 20 examines the predicted signals of the local $E_z$ values at several random positions within the domain from the multiscale model that consists of sub-models with step sizes $TS$ = 1 – 64. Additionally, Figure 21 displays predicted 2D spatial snapshots of the forecasted and the true $E_z$ fields at sample time instants, demonstrating the model's ability to retain spatial coherence.

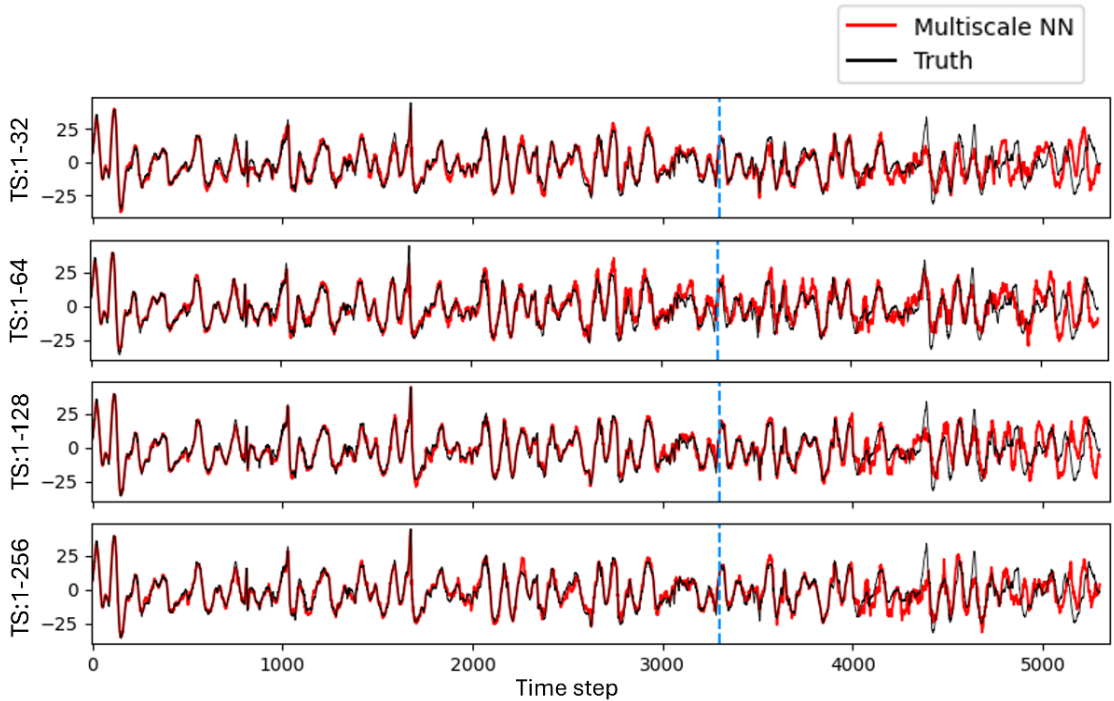

Figure 19: Reconstruction and forecasts from several multiscale networks for the axial-azimuthal E × B plasma test case: time evolutions of the spatially averaged azimuthal electric field ($E_z$). Each row represents a multiscale architecture with different number of sub-models with step sizes TS = 1, 2, …, $TS_{max}$, where $TS_{max}$ denotes the sub-model with the maximum step size in that row. The vertical dashed blue lines separate the training and test intervals.

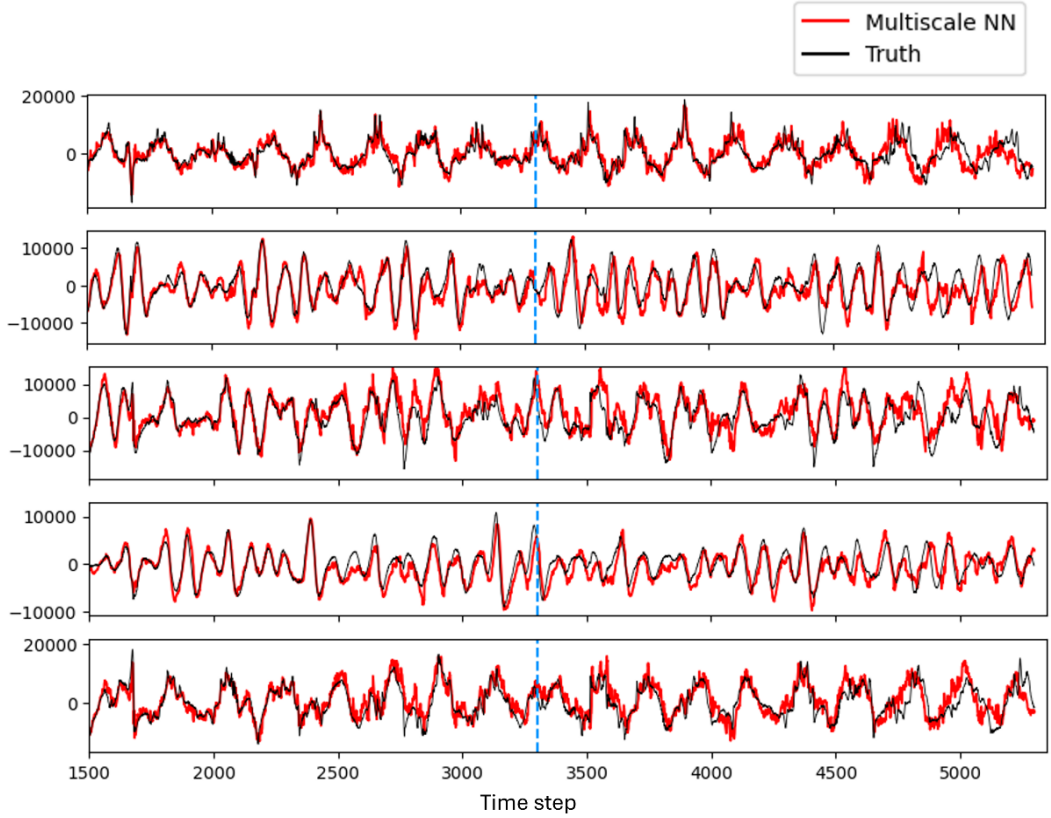

Figure 20: Reconstruction and forecasts from the multiscale network with sub-models of $TS = 1, 2, \ldots, 64$ for the axial-azimuthal E × B plasma case: time evolutions of the local values of azimuthal electric field ($E_z$) at different random locations with the domain, each represented in one row. The dashed blue lines separate the training and test intervals. Note that, for better clarity, only a portion of the training sequence is displayed in the figure, omitting data points before time step 1500.

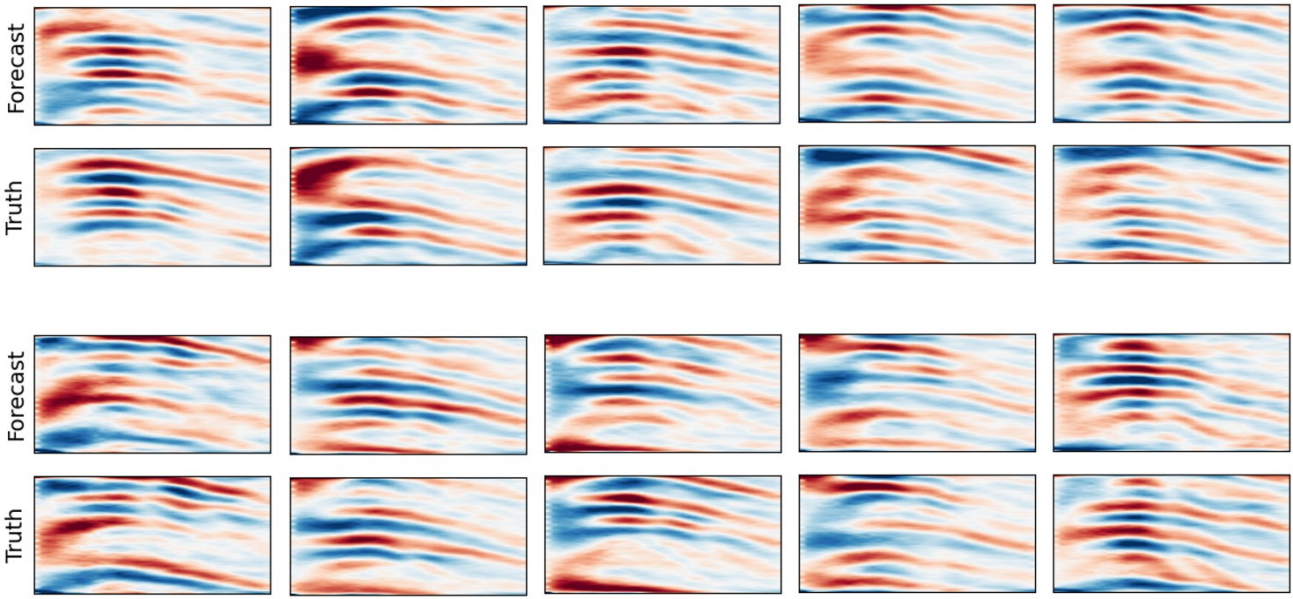

Figure 21: Forecasted 2D snapshots of the azimuthal electric field ($E_z$) from the multiscale model consisting of sub-models with step sizes $TS = 1, 2, \ldots, 64$ at various sample time instants within the test interval compared against the corresponding ground-truth snapshots. In all plots, the horizontal axis represents the axial coordinate ($x$), and the vertical axis represents the azimuthal coordinate ($z$).

Finally, Figure 22 evaluates the forecasting performance of the individual sub-models trained with different step sizes, indicating a step-size-dependent error accumulation behavior that is consistent with the observations from the previous test cases. Small step-size sub-models ($TS \leq 16$) show rapid error accumulation, whereas larger step-size sub-models remain stable for a longer period. Once again, the results highlight the efficacy of the multiscale architecture in mitigating errors by combining the resolution of the short-time-scale sub-models with the stability of the large-time-scale model components.

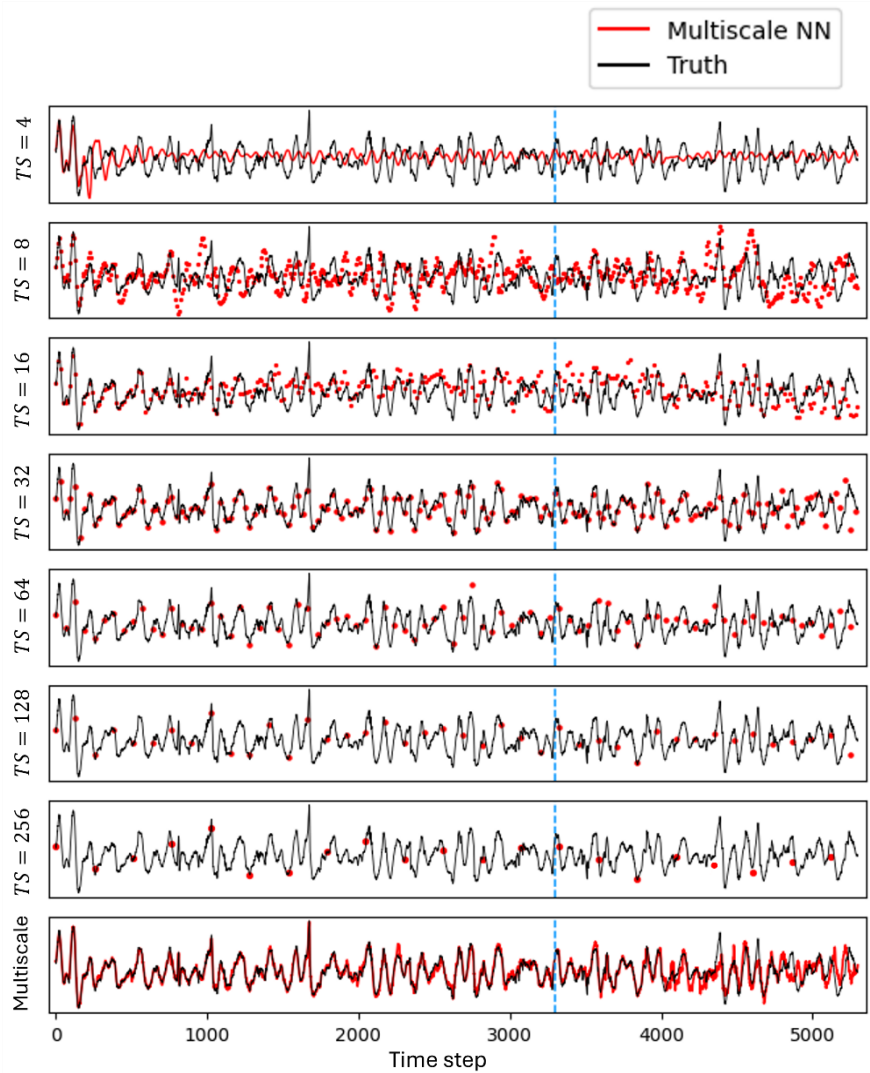

Figure 22: Time evolution of the spatially averaged azimuthal electric field ($E_z$) from individual sub-models with different step sizes and from the multiscale model (bottom-most row) consisting of models with step sizes TS = 1, 2, …, 256 compared against ground-truth data for axial-azimuthal E × B plasma case. The dashed blue lines separate the training and test intervals.

**Section 5: Conclusion**

In this effort, we presented the application of a hierarchical multiscale neural network architecture for autonomous forecasting of plasma dynamics. By integrating multiple time-stepping neural network models operating at different temporal resolutions, the multiscale approach successfully mitigates the common challenges associated with single-timescale ML recursive forecasting, such as error accumulation and numerical instability. The results from canonical dynamical systems, namely, the Van der Pol oscillator, the Cubic Oscillator, and the Lorenz system, demonstrated the advantages of the multiscale approach in maintaining predictive accuracy over long time horizons compared to single-timescale architectures.

The effectiveness of the multiscale framework was further demonstrated through its application to E × B plasma systems of real-world applied relevance. The multiscale model preserved key dynamical features across scales, outperforming single-timestep neural networks that struggled with compounding errors for autonomous forecasting. The balance between fine-resolution short-timestep models and stable long-timestep models allowed for improved predictiveness, underlining the suitability of the hierarchical method for multiscale physical systems.

While the multiscale model extends the prediction horizon beyond that of conventional ML architectures, its forecasting capability remains inherently limited by the fundamental extrapolation and generalizability constraints of data-driven methods. Like other neural-network-based approaches, it ultimately relies on patterns learned from the training data, due to which highly transient regimes and overly complex behaviors can still lead to forecast degradation.

Although in this work we used ResNets with LSTM layers, the multiscale architecture serves as a flexible framework, allowing the integration of various ML architectures as sub-models. By employing more advanced or domain-specific ML architectures within this hierarchical framework, the prediction horizon and generalizability may be improved.

The method's parallelizability and multiscale structure make it particularly well-suited as the underlying framework for plasma systems' digital twins [58][59]. Plasma dynamics inherently evolve across a wide range of temporal scales, from fast fluctuations to slow, long-term variations. The hierarchical multiscale architecture effectively decomposes the system into specialized sub-models, each dedicated to capturing a specific time scale at an optimal resolution.

Importantly, the multiscale structure of the model enables efficient parallel computing, where different temporal components can be solved simultaneously rather than sequentially, which significantly accelerates computation. This parallelizability is especially important for real-time monitoring and control applications using digital twins of plasma technologies [58][59].

**Appendix A: Dataset generation and neural network setup for the canonical dynamic systems**

The models for the Van der Pol, the Cubic Oscillator and the Lorenz dynamical systems are trained on a single long trajectory consisting of 1,000,000 data points. The trajectories are simulated by numerically integrating the system of ordinary differential equations (ODEs), describing the respective dynamical systems, using the "solve_ivp" function, which is part of the SciPy library [60] (in the "scipy.integrate" package) and is used to solve initial value problems (IVPs). The integrator was part of the open-source code by Liu et al. [47]. The integration time step for generating data for the Lorenz system is 1 ms and 10 ms for the Van der Pol system and the Cubic Oscillator.

The canonical dynamical systems are represented by coupled nonlinear ODEs as listed in Table 3. In these equations, $x$, $y$ and $z$ serve as state variables, capturing the system's evolution over time ($t$).

Table 3 also presents the architecture of the neural network components in the multiscale model. In the table's third column from left, the numbers indicate the size of each layer: the first and last numbers correspond to the input and output sizes, which match the number of state variables, while the middle numbers represent the hidden layer sizes for fully connected layers or the hidden state size for LSTM layers. The last column from left provides the step size (represented by $TS$ in the caption of the relevant figures in the body of the text) for each neural network component.

|  | **System of equations** | **NN components architecture** | **Step size of NN components** |
| --- | --- | --- | --- |
| **Van der Pol system** | $\dot{x} = y,$ $\dot{y} = 2(1-x^2)y - x.$ | [2, 256, 256, 256, 2] | $1, 2, 4, \ldots, 2^{12}$ |
| **Cubic Oscillator** | $\dot{x} = -0.1x^3 + 2y^3,$ $\dot{y} = -2x^3 - 0.1y^3.$ | [2, 256, 256, 256, 2] | $1, 2, 4, \ldots, 2^{12}$ |
| **Lorenz attractor** | $\dot{x} = 10(y-x),$ $\dot{y} = 28x - xz - y,$ $\dot{z} = xy - \frac{8}{3}z.$ | [3, 256, 256, 256, 3] | $1, 2, 4, \ldots, 2^{12}$ |

Table 3: Summary of the canonical dynamical systems used for benchmarking, detailing the corresponding system of ODEs and the neural network architectures employed to model them.

**Appendix B: Supplementary results for the radial-azimuthal $E \times B$ plasma test case**

This appendix includes additional results supporting the discussions in subsection 4.2.1 on the performance of the multiscale model in forecasting various plasma properties in the radial-azimuthal $E \times B$ configuration.

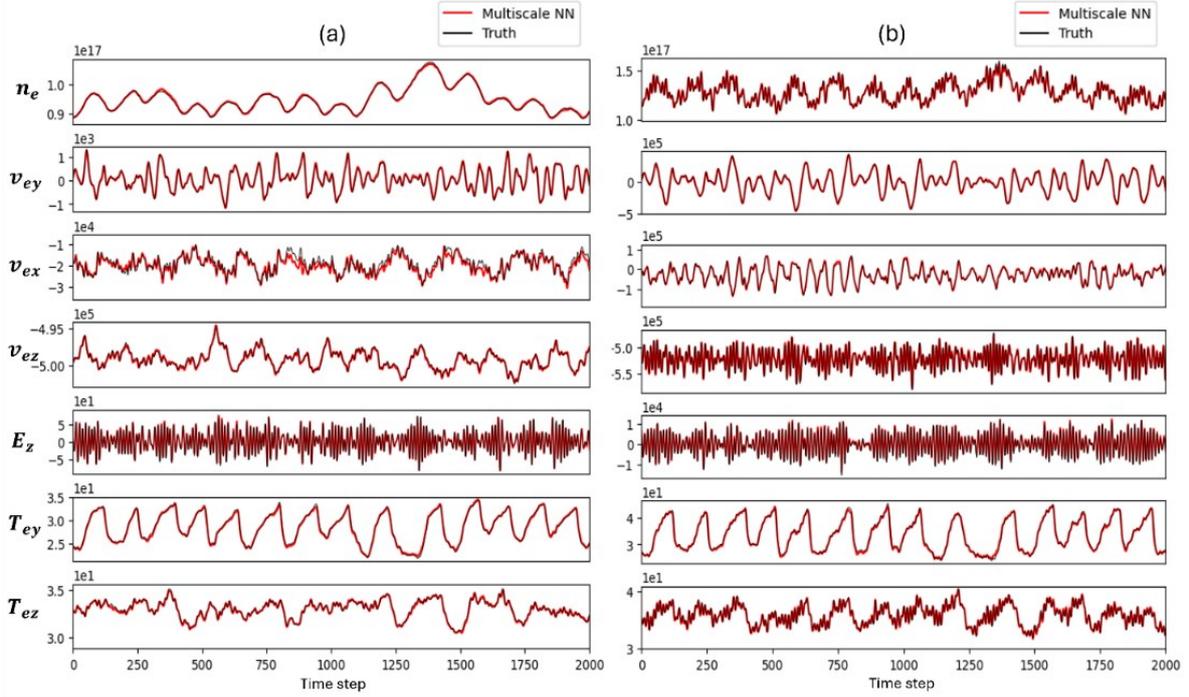

Figure 23: Reconstruction of the training data from the multiscale network consisting of models with step sizes $TS$ = 1, 2, …, 64, trained using 3-step-ahead training strategy for the radial-azimuthal E × B plasma test case: time evolution of (**a**) spatial average and (**b**) mid-domain local values of different plasma properties, namely, electron number density ($n_e$), electrons' axial, radial, and azimuthal velocities ($v_{ex}$, $v_{ey}$ and $v_{ez}$ respectively), azimuthal electric field ($E_z$), and radial and azimuthal electron temperatures ($T_{ey}$ and $T_{ez}$, respectively).

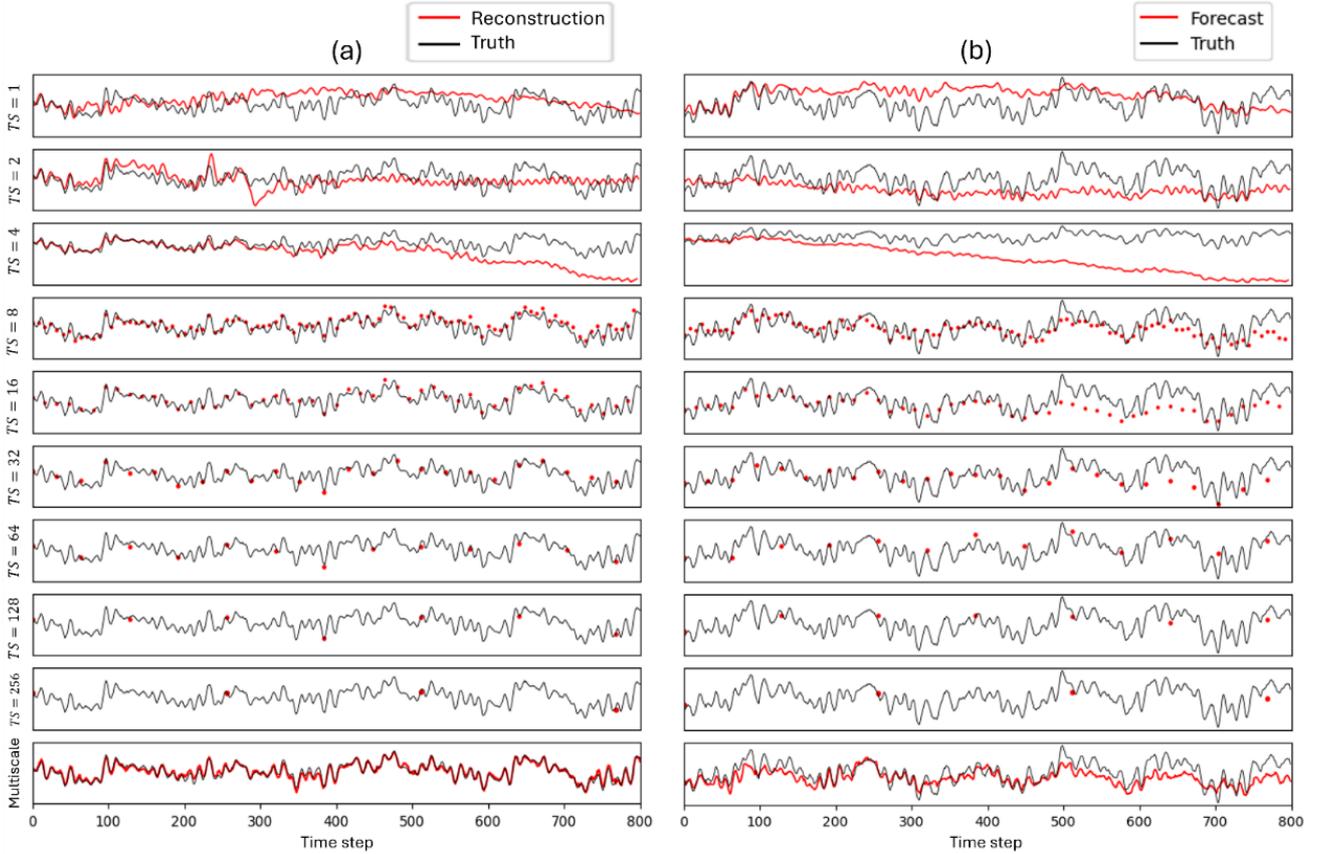

Figure 24: Time evolutions of the spatially averaged electrons' axial drift velocity ($v_{ex}$) for the radial-azimuthal E × B plasma test case: (**a**) reconstructions of the training data, and (**b**) forecasts of the test data from the individual sub-models with different step sizes and from the multiscale model (bottom-most row) comprising models with step sizes $TS$ = 1, 2, …, 256 compared against ground-truth data.

Figure 23 presents the reconstructions of the training data, showing the time evolution of both the spatially averaged values and the mid-domain local values for key output plasma properties. These plots serve as pre-assessment to ensure the trained model's ability to reproduce the training sequence before it being applied for forecasting beyond the training window.

Figure 24 focuses on reconstruction of the training sequence and forecasting during the test interval using individual sub-models with various step sizes for an important plasma property, the electrons' axial drift velocity ($v_{ex}$). This plasma property quantifies the cross-magnetic-field transport of the electrons – a key unresolved question in plasma physics. The results are shown in terms of the temporal evolution of the spatially averaged $v_{ex}$ values, indicating the impact of NN model step size on error accumulation and, consequently, the predictive accuracy.

Figure 25 extends the analysis in Figure 24 to multiple plasma properties, illustrating the time variations of the mid-domain local values.

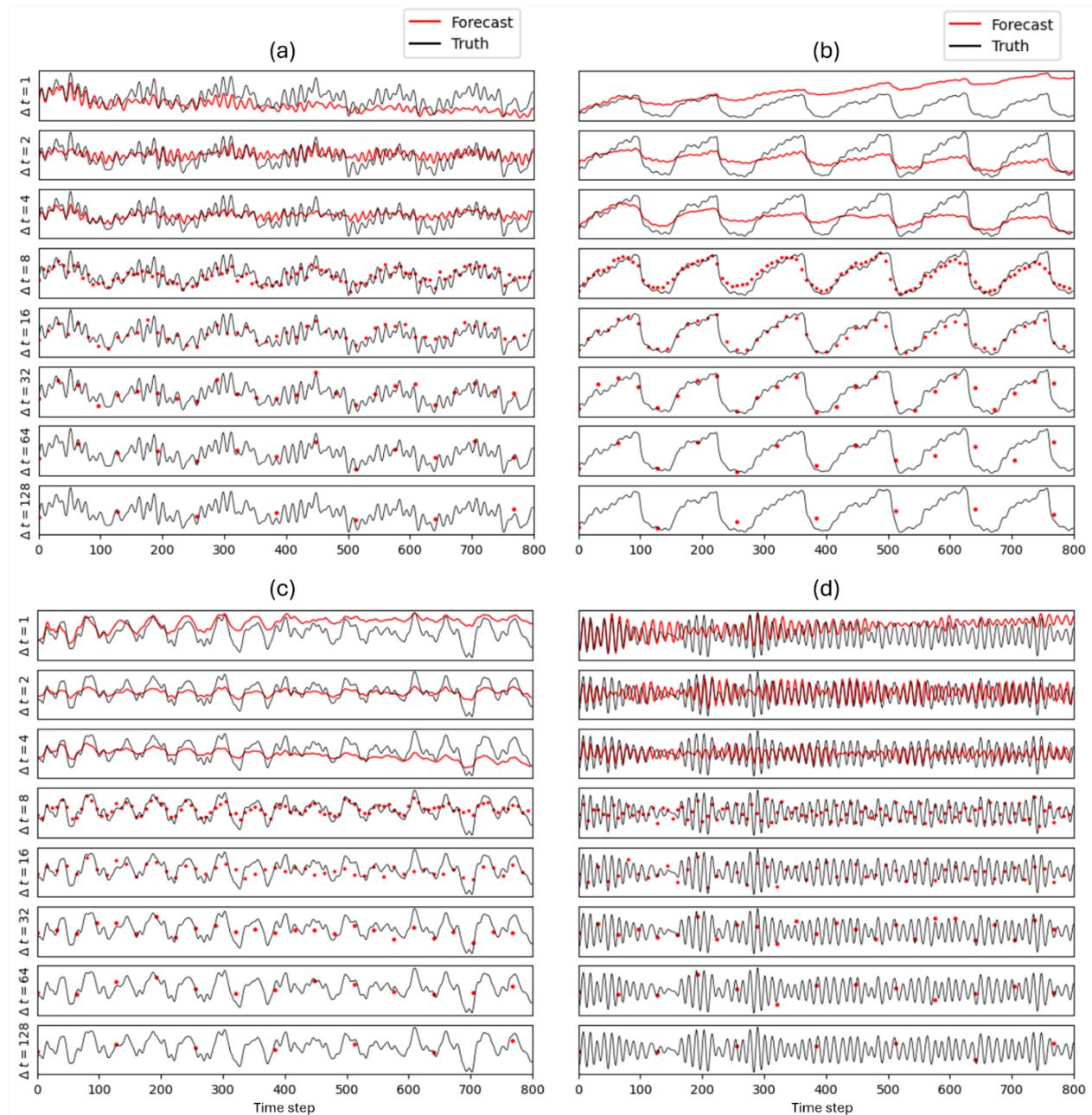

Figure 25: Forecasts of several plasma properties from individual sub-models with different step sizes for the radial-azimuthal E × B plasma test case: time evolutions of the mid-domain local values of (**a**) electron number density ($n_e$), (**b**) radial electron temperature ($T_{ey}$), (**c**) electrons' axial velocity ($v_{ex}$), and (**d**) the azimuthal electric field ($E_z$).

**Acknowledgments**:

There was no funding supporting this activity.

**Data Availability Statement**:

The data that support the findings of this study are available from the corresponding author upon reasonable request.